\documentclass[runningheads]{llncs}
\usepackage{graphicx}
\usepackage[subtle]{savetrees}

\usepackage{url}
\usepackage{wrapfig}
\usepackage{amssymb}
\usepackage{algorithm, algorithmic}
\usepackage{amsmath}
\usepackage{subfig}
\usepackage{xcolor}
\usepackage[style=base]{caption}

\makeatletter
\newcommand{\removelatexerror}{\let\@latex@error\@gobble}
\makeatother
\makeatletter
\DeclareRobustCommand{\circ}{\mathord{\mathpalette\is@circle\relax}}
\newcommand\is@circle[2]{%
  \begingroup
  \sbox\z@{\raisebox{\depth}{$\m@th#1\bigcirc$}}%
  \sbox\tw@{$#1\square$}%
  \resizebox{!}{\ht\tw@}{\usebox{\z@}}%
  \endgroup
}
\makeatother
\newcommand\mc{\mathcal}

\newtheorem{definition2}{Definition}
\newtheorem{assumption}{Assumption}
\newtheorem{problem2}{Problem}
\begin{document}
%
\title{Towards Better Test Coverage: Merging Unit Tests for Autonomous Systems$^{\dagger}$}
%
%
\author{Josefine B. Graebener\inst{*} \qquad
Apurva Badithela\inst{*} \qquad
Richard M. Murray}
\authorrunning{J. B. Graebener, A. Badithela, R. M. Murray}
\titlerunning{Towards Better Test Coverage: Merging Unit Tests for Autonomous Systems}
%
\institute{California Institute of Technology, Pasadena CA 91125, USA
\\\email{\{jgraeben, apurva, murray\}@caltech.edu}}
\maketitle              
\def\thefootnote{*}\footnotetext{These authors contributed equally to this work.}\def\thefootnote{\arabic{footnote}}
\def\thefootnote{$^{\dagger}$}\footnotetext{The code for examples given in this paper can be found at:  \url{https://github.com/jgraeb/MergeUnitTests}.}

\begin{abstract}
We present a framework for merging unit tests for autonomous systems. Typically, it is intractable to test an autonomous system for every scenario in its operating environment. The question of whether it is possible to design a single test for multiple requirements of the system motivates this work. First, we formally define three attributes of a test: a test specification that characterizes behaviors observed in a test execution, a test environment, and a test policy. Using the merge operator from contract-based design theory, we provide a formalism to construct a merged test specification from two unit test specifications. Temporal constraints on the merged test specification guarantee that non-trivial satisfaction of both unit test specifications is necessary for a successful merged test execution. We assume that the test environment remains the same across the unit tests and the merged test. Given a test specification and a test environment, we synthesize a test policy filter using a receding horizon approach, and use the test policy filter to guide a search procedure (e.g. Monte-Carlo Tree Search) to find a test policy that is guaranteed to satisfy the test specification. This search procedure finds a test policy that maximizes a pre-defined robustness metric for the test while the filter guarantees a test policy for satisfying the test specification. We prove that our algorithm is sound. Furthermore, the receding horizon approach to synthesizing the filter ensures that our algorithm is scalable. Finally, we show that merging unit tests is impactful for designing efficient test campaigns to achieve similar levels of coverage in fewer test executions. We illustrate our framework on two self-driving examples in a discrete-state setting.

\end{abstract}
\section{Introduction}
\label{sec:intro} 
Rigorous test and evaluation of autonomous systems is imperative for deploying autonomy in safety-critical settings~\cite{seshia2016towards}. In the case of testing self-driving cars, operational tests are constructed manually by experienced test engineers and can be combined with test cases generated in simulators using falsification techniques~\cite{fremont2020formal}. In addition, operational testing of self-driving cars on the road is expensive, and would need to be  repeated after every design iteration~\cite{kalra2016driving}. In this paper, we pose the question of whether it is possible to check multiple requirements in a single test execution. Addressing this question is the first step towards optimizing for the largest number of test requirements checked in as few operational tests as possible.

The study of principled approaches to testing, verification and validation is a relatively young but growing research area. In the formal methods community, \emph{falsification} is the technical term referring to the study of optimization algorithms, typically black-box, and sampling techniques to search for inputs that result in the system-under-test violating its formal requirements on input-output behavior~\cite{abbas2013probabilistic,annpureddy2011s,dreossi2019compositional,dreossi2019verifai,ghosh2018verifying,sankaranarayanan2012falsification}. Falsification algorithms require a metric defined over temporal logic requirements to quantitatively reason about the degree to which a formal requirement has been satisfied. Assuming that the design of the autonomous system is black-box, falsification algorithms seek to find inputs that minimize the metric associated with satisfying formal requirements. The reasoning here is that minimizing this metric brings the system closer to violating its \emph{requirements}, thus being a critical test scenario~\cite{klischat2020scenario}. 
Formal methods literature uses \emph{falsification} and \emph{testing} interchangeably. In addition to manually constructing operational tests, falsification is used to find critical scenarios in simulation and the test environment parameters characterizing these critical scenarios are  used for operational testing~\cite{fremont2020formal}.  
\begin{figure}[!b]
\vspace{-5mm}
\centering
\includegraphics[width=\textwidth]{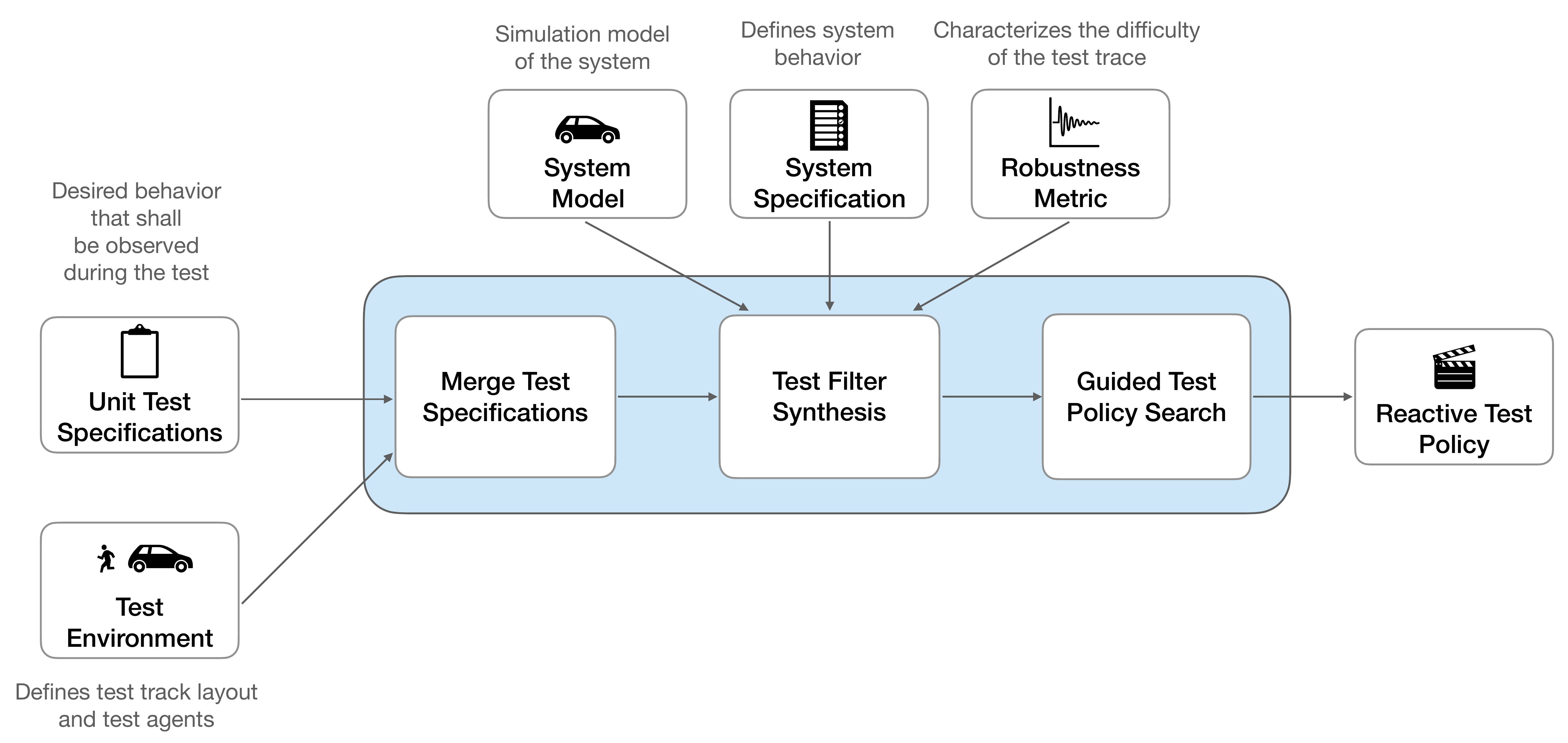}
\caption{Overview of the proposed framework. The blocks the left represent the inputs to the algorithm that define the unit tests, the blocks on the top represent inputs describing the system under test, the building blocks of our approach are shown in the blue shaded box, and the test policy is the result of the algorithm.} \label{fig:framework}
\end{figure}
Falsification aims to find parameters of the test environment that lead the system to violate its \emph{requirements}. However, our approach is different in that we construct a test with respect to a test specification, which characterizes a set of desired test executions. For example, consider an autonomous car on a test track. The requirement for the autonomous car is to drive around the track and follow traffic rules while the human drivers of the test vehicles are instructed to drive in a specific fashion (ex: maintaining some distance between each other). These guidelines given to the test drivers constitute the test specification, which is not known to the system-under-test. Instead of considering all possible test environment policies, the test specification restricts the space of scenarios that our test policy search algorithm searches over. It also leverages reactivity: test scenarios are not planned in advance, but the test environment agents will react depending on the actions taken by the system under test. 

Our contributions are the following. First, we formally characterize a test by three attributes, a test specification, a test environment and a test policy. Second, we leverage the merge operator from assume-guarantee contract theory to merge two unit test specifications into a merged specification, resulting in a single test that checks the test specifications of both unit tests. Furthermore, if necessary, we characterize temporal constraints on the merged test specification. Finally, we use Monte Carlo Tree Search (MCTS) to search for a test environment policy corresponding to the test specification, and use receding horizon synthesis techniques to prevent the search procedure from exploring policies that violate the test specification. This framework is illustrated in Figure~\ref{fig:framework}.


\vspace{-3mm}
\section{Background}
\label{sec:backg}
\vspace{-2mm}
In this work, we choose Linear Temporal Logic (LTL) to represent the system and test specifications. LTL is a temporal logic language for describing linear-time properties over traces of computer programs and formally verifying their properties~\cite{pnueli1977temporal}. Although first introduced to formally describe properties of computer programs, LTL has been used for formal methods applications in control such as temporal logic synthesis of planners and controllers~\cite{wongpiromsarn2012receding,kress2009temporal,kloetzer2008fully}.
\begin{definition2}[Linear Temporal Logic (LTL)~\cite{baier2008principles}]
Given a set of atomic propositions \(AP\), the \emph{syntax} of LTL is given by the following grammar:
\vspace{-2mm}
\begin{equation}
    \varphi ::= \text{true}\:|\:a\:|\:\varphi_1 \land \varphi_2\:|\:\neg \varphi \:|\: \circ \varphi\:|\: \varphi_1 \mathcal{U} \varphi_2\,
\vspace{-2mm}
\end{equation}
    where $a \in AP$ is an atomic proposition, \(\wedge\) (and) and \(\neg\) (not) are logic operators, and $\circ$ (next) and $\mathcal{U}$ (until) are temporal operators. Other temporal operators such as $\square$ (always), $\lozenge$ (eventually), \(\square \lozenge\) (always eventually), and \(\lozenge \square\) (eventually always) can be derived.
    Let \(\varphi\) be an LTL formula over the set of atomic propositions \(AP\). The \emph{semantics} of LTL are inductively defined over an infinite sequence of states \(\sigma = s_0s_1s_2\ldots\) as follows: i) If \(p \in AP\), \(s_i \models p\) iff \(p\) evaluates to true at \(s_i\), ii) \(s_i \models \neg \varphi\) iff \(s_i \not\models \varphi\), iii) \(s_i \models \varphi_1 \wedge \varphi_2\) iff  \(s_i \models \varphi_1 \wedge s_i \models \varphi_2\), iv)  \(s_i \models \circ \varphi\) iff \(s_{i+1} \models \varphi\), v) \(s_i \models \varphi_1 \mathcal{U} \varphi_2\) iff \(\exists j > i\) such that for all \(k \in [i,j)\), \(s_k \models \varphi_1\) and \(s_j \models \varphi_2\). An infinite sequence \(\sigma = s_0s_1\ldots\) satisfies an LTL formula \(\varphi\), denoted by \(\sigma \models \varphi\), iff \(s_0 \models \varphi\).
\end{definition2}

In our framework, we consider a fragment of LTL specifications in the class of generalized reactivity of rank 1 ($GR(1)$)~\cite{piterman2006synthesis}. $GR(1)$ specifications are expressive for capturing safety (\(\square\)), liveness (\(\lozenge\)), and recurrence (\(\square \lozenge\)) requirements that are relevant to several autonomous systems~\cite{kress2009temporal,wongpiromsarn2012receding}. A \(GR(1)\) formula \(\varphi\) is as follows,
\begin{equation}
    \varphi = (\varphi^{\text{init}}_e \land \square \varphi^s_e \land \square \lozenge \varphi^f_e) \rightarrow (\varphi^\text{init}_s\land  \square \varphi^s_s \land \square \lozenge \varphi^f_s) \, ,
\label{eq:orig_gr1}
\end{equation}
where the subscript $s$ refers to the robotic system for which a reactive controller is being synthesized, and \(\varphi_{s}^{\text{init}}\), \(\square \varphi_{s}^s \), and \(\square \lozenge \varphi_{s}^f\), define respectively, the initial requirements, safety requirements and recurrence requirements on the system denoted by $s$. Similarly, $\varphi_{e}^{\text{init}}$, \(\square \varphi_{e}^s \), and \(\square \lozenge \varphi_{e}^f\), define requirements on the environment $e$ of the system $s$. Furthermore, synthesis for $GR(1)$ formulas has time complexity \(O(|V|^3)\), where \(|V|\) is the size of the state space~\cite{piterman2006synthesis}.
\vspace{-2mm}
\subsubsection{Assume-Guarantee Contracts}
Contract-based design was first developed as a formal modular design methodology for analysis of  component-based software systems~\cite{meyer1992applying,dijkstra1975guarded,lamport1990win}, and later applied for the design and analysis of complex autonomous systems~\cite{nuzzo2013contract,filippidis2018layering}. 
In this work, we adopt the mathematical framework of assume-guarantee contracts presented in~\cite{benveniste2018contracts,passerone2019coherent}. 
\begin{definition2}[Assume-Guarantee Contract]
\label{def:contracts}
Let $\Lambda$ be an alphabet and $\mathcal{B}(\Lambda)$ be the set of all behaviors over $\Lambda$. A component $M$ over the alphabet $\Lambda$ is defined as $M\subseteq B(\Lambda)$. Then an \textit{assume-guarantee contract} $\mathcal{C}$ is defined as a pair $\mathcal{C}=(A,G)$, where $A$ is a set of behaviors for assumptions on the environment in which the component operates, and $G$ is a set of behaviors for the guarantees that the component provides, assuming its assumptions on the environment are met. $M$ is an implementation of a contract, $M\models\mathcal{C}$, if and only if $M\subseteq G\: \land \:\neg A \:\Leftrightarrow\: M \land (A \land \neg G) = \emptyset$  \cite{benveniste2007multiple}.
\end{definition2}

In this work, the assumptions and guarantees constituting assume-guarantee contracts are LTL formulas. To facilitate the contract algebra, we will consider contracts in their \textit{saturated} form, where a contract is defined as $\mathcal{C}=(A,A\rightarrow G)$. In Section~\ref{sec:prob_setup} we define system and test specifications with LTL and borrow operators from assume-guarantee contract theory in Section~\ref{sec:merge_spec}  to formally define the merge of two unit tests. 
\vspace{-3mm}
\section{Problem Setup}
\label{sec:prob_setup}
\vspace{-1mm}
First we define the system under test, which we will refer to as \textit{system} for brevity, and its corresponding system specification. We assume that the system, the system specification, and the controller are provided by the designer of the system and cannot be modified when designing the test.
\begin{definition2}[Transition System]
A \textit{transition system} is a tuple $\mathcal{T}:=(Q,\rightarrow)$, where $Q$ is a set of states and $\rightarrow \: \subseteq Q \times Q$ is a transition relation. If $\exists$ a transition from $q_1 \in Q$ to $q_2 \in Q$, we write $q_1 \rightarrow q_2$.
\end{definition2}
\begin{definition2}[System]
Let $\mathbb{V}_S$ be the set of system variables, and let \(Q_\text{sys}\) be the set of all possible valuations of \(\mathbb{V}_\text{sys}\). A \textit{system} $S$ is a transition system $\mc{T}_\text{sys} = (Q_\text{sys},\rightarrow_\text{sys})$, where the transition relation $\rightarrow_\text{sys}$ is defined by the dynamics of the system. 
\end{definition2}
\begin{definition2}[System Specification]
A \textit{system specification} \(\varphi_\text{sys}\) is the $GR(1)$ formula,
\vspace{-2mm}
\begin{equation}
    \varphi_\text{sys} = (\varphi^{init}_{\text{test}} \land \square \varphi^s_\text{test} \land \square \lozenge \varphi^f_{\text{test}}) \rightarrow (\varphi^\text{init}_\text{sys}\land  \square \varphi^s_{\text{sys}} \land \square \lozenge \varphi^f_{\text{sys}})\, ,
    \label{eq:sys_spec}
\end{equation}
where \(\varphi^\text{init}_{\text{sys}}\) is the initial condition that the system needs to satisfy,  \(\varphi^s_{\text{sys}}\) encode system dynamics and safety requirements on the system, and \(\varphi^f_{\text{sys}}\) specifies recurrence goals for the system. Likewise, \(\varphi^\text{init}_{\text{test}}\), \(\varphi^s_{\text{test}}\), and \(\varphi^f_{\text{test}}\) represent assumptions the system has on the test environment.
\end{definition2}


The system is evaluated in a test environment, which comprises of both the test track and test agents. A \emph{test} is characterized by the \emph{test environment}, a \emph{test specification}, and a \emph{test policy}. Our approach differs from falsification in that we are not generating a test strategy to stress test the system for \(\varphi^\text{init}_\text{sys}\land  \square \varphi^s_{\text{sys}} \land \square \lozenge \varphi^f_{\text{sys}}\). Instead, we synthesize a test for a new concept --- a test specification --- which describes the set of behaviors we would like to see in a test. For example, an informal version of a test specification is requiring test agents to ``drive around the test track at a fixed speed while maintaining a certain distance from each other". 
\begin{definition2}[Test Environment]
Let $\mathbb{V}_\text{test}$ be the set of test environment variables, and let \(Q_\text{test}\) be the set of all possible valuations of \(\mathbb{V}_\text{test}\). A \textit{test environment} $T$ is a transition system $\mc{T}_\text{test} = (Q_\text{test},\rightarrow_\text{test})$, where the transition relation $\rightarrow_\text{test}$ is defined by the dynamics of the test agents. 
\end{definition2}
\begin{definition2}[Test Specification]
A \textit{test specification} \(\varphi_\text{test}\) is the $GR(1)$ formula,
\vspace{-2mm}
\begin{equation}
    \varphi_\text{test} := \big(\varphi^{init}_\text{sys} \wedge \square \varphi^{s}_\text{sys} \wedge \square \lozenge \varphi^{f}_\text{sys}\big) \rightarrow \big(\varphi^{init}_\text{test} \wedge \square \varphi^{s}_\text{test} \wedge \square \lozenge \varphi^{f}_\text{test} \wedge \square\psi^{s}_\text{test} \wedge \square \lozenge \psi^{f}_\text{test} \big)\, ,
    \vspace{-2mm}
    \label{eq:test_spec}
\end{equation}
where \(\varphi^{init}_\text{sys} \), \(\varphi^{s}_\text{sys}\) and \(\varphi^{f}_\text{sys}\), \(\varphi^{init}_\text{test}\), \(\varphi^{s}_\text{test}\) and  \(\varphi^{f}_\text{test}\) are propositional formulas from equation~\eqref{eq:sys_spec}. Additionally, \(\square \psi^{s}_\text{test}\) and \(\square \lozenge \psi^{f}_\text{test}\) describe the safety and recurrence formulas for the test environment in addition to the dynamics of the test environment known to the system. Note that the system is unaware of these additional specifications on the test environment, and the test specification is such that the system is allowed to satisfy its requirements. Defining the test specification in this manner allows for i) synthesizing a test in which the system, if properly designed, can meet \(\varphi_\text{sys}\), and ii) specifying additional requirements on the test environment, unknown to the system at design time. We assume that test specifications are defined \emph{a priori}; we leave finding relevant test specifications to future work.
\end{definition2}


Let \(\mc{T}_\text{prod} = (Q_\text{prod}, \rightarrow_\text{prod})\) be a turn-based product transition system constructed from \(\mc{T}_\text{sys}\) and \(\mc{T}_\text{test}\), where \(Q_\text{prod} := Q_\text{sys} \times Q_\text{test}\), and \(\rightarrow_\text{prod} \subseteq Q_\text{prod} \times Q_\text{prod}\). In particular, for every transition \((s,s') \in \rightarrow_\text{sys}\), we have \(((s,t),(s',t)) \in \rightarrow_\text{prod}\) where \(t \in Q_\text{test}\). Similarly, for every transition \((t,t') \in \rightarrow_\text{test}\), we have \(((s,t),(s,t')) \in \rightarrow_\text{prod}\) where \(s \in Q_\text{sys}\).

\begin{definition2}[Game Graph]
\label{def:game_graph}
Let \(V_\text{sys}\) and \(V_\text{test}\) be copies of the states \(Q_\text{prod}\). Let \(E_{sys}\) denote the set of transitions \(((s,t), (s',t)) \in \rightarrow_\text{prod}\), and let \(E_{\text{test}}\) denote the set of transitions \(((s,t), (s,t')) \in \rightarrow_\text{prod}\) for some \(s,s' \in Q_\text{sys}\) and \(t, t' \in Q_\text{test}\). Then the \textit{game graph} $G = (V,E)$ is a directed graph with vertices $V :=V_\text{sys} \cup V_\text{test}$ and edges $E :=E_\text{sys} \cup E_\text{test}$.
\end{definition2}


\begin{definition2}[Policy]
On the game graph \(G\), a policy for the system is a function \(\pi_\text{sys}: V^* V_\text{sys} \rightarrow V_\text{test}\) such that \((s, \pi_\text{sys}(w\cdot s)) \in E_\text{sys}\), where \(s \in V_\text{sys}\) and \(w \in V^*\). Similarly defined, \(\pi_\text{test}\) denotes the test environment policy, where $^*$ is the Kleene star operator.
\end{definition2}

\begin{definition2}[Test Execution]
A \textit{test execution} $\sigma = v_0 v_1 v_2 \ldots$  starting from vertex \(v_0 \in V\) is an infinite sequence of states on the game graph $G$. Since \(G\) is a turn-based game graph, the states in the test execution alternate between \(V_\text{sys}\) and \(V_\text{test}\), so if \(V_{1} \in V_\text{sys}\), then \(v_{i+1} = \pi_\text{sys}(v_0\ldots V_{1})\). Let \(\sigma_{\pi_\text{sys}\times \pi_\text{test}}(s_0)\) be the test execution starting from state \(s_0 \in V_\text{sys}\) for policies \(\pi_\text{sys}\) and \(\pi_\text{test}\). Let \(\Sigma\) denote the set of all possible test executions on \(G\). A robustness metric \(\rho: \Sigma \rightarrow \mathbb{R}\) is a function evaluated assigning a scalar value to a test execution. 
\end{definition2}
\begin{problem2}
\label{problem_statement}
Given system and environment transition systems, \(\mc{T}_\text{sys}\) and \(\mc{T}_\text{test}\), two unit test specifications $\varphi_{\text{test}, 1}$ and $\varphi_{\text{test}, 2}$, and a robustness metric $\rho$, find a test policy $\pi^*_\text{test}$, such that 
\vspace{-2mm}
\begin{equation}
\label{eq:orig_problem_statement}
\begin{aligned}
\pi^*_\text{test}\quad = \quad \arg \, \max_{\pi_\text{test}} \quad & \rho(\sigma_{\pi_\text{sys} \times \pi_\text{test}})\\
\textrm{s.t.} \quad &  \sigma_{\pi_\text{sys} \times \pi_\text{test}} \models (\varphi_{\text{test},1}
 \land \varphi_{\text{test},2}) \,, \quad \forall \: \pi_\text{sys} \models \varphi_\text{sys},\\
\end{aligned}
\vspace{-2mm}
\end{equation}
\end{problem2}
\vspace{-2mm}
\subsubsection{Running Example --- Lane Change}
\begin{figure}
\begin{minipage}[c]{0.55\textwidth}
\vspace{-2mm}
\includegraphics[width=0.95\textwidth]{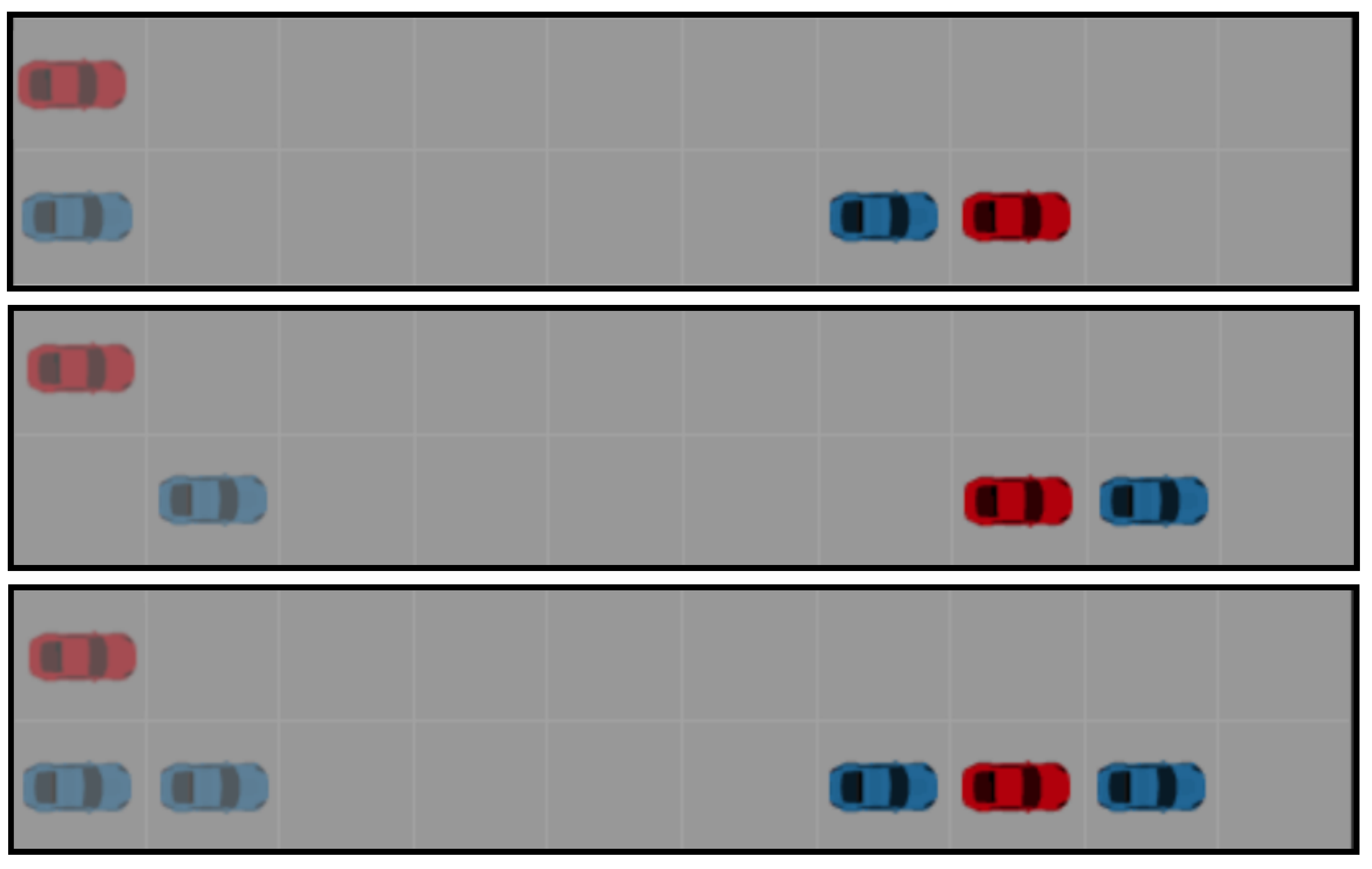}
\end{minipage}\hfill
\begin{minipage}[c]{0.45\textwidth}
\caption{Lane change example with initial (left) and final (right) configurations. The \(x\)-coordinates are numbered from left to right, and \(y\)-coordinates are numbered top to bottom, starting from 1. The system (red) is required to merge into the lower lane without colliding. Merging in front of (top), behind (center), or in between (bottom) tester agents (blue). } \label{fig:merge_ex}
\end{minipage}
\vspace{-5mm}
\end{figure}

Consider the example of lane change illustrated in Figure~\ref{fig:merge_ex}. The system (red car) must merge into the lower lane before the track ends, and must not collide with the test environment agents (blue cars). Thus, the liveness requirement of changing lanes, \(\varphi^f_\text{sys} := (y_\text{sys} = 2)\), and the safety requirement of not colliding with test agent \(i\), \(\neg(y_\text{sys} = y_{\text{test},i} \wedge x_\text{sys} = x_{\text{test},i}) \in \varphi^s_\text{sys}\), constitute part of the system specification \(\varphi_\text{sys}\). In the two unit tests, we have the system changing into the other lane in front of and behind a tester car, respectively, and in the merged test, it finished its lane change maneuver in between the tester cars.
\vspace{-5mm}
\label{sec:ex_lane_change}

\vspace{-4mm}
\section{Merging Unit Tests}
\label{sec:approach}
\vspace{-2mm}
In this section, we will outline our main approach for merging unit tests. First, we define the notion of a merged test and use the merge operator for merging test specifications and add temporal constraints to the test specification, if necessary. Then, we construct an auxiliary graph corresponding to the merged test specification and describe the synthesis of the test policy filter on this auxiliary graph using a receding horizon approach.
\subsection{Merging Test Specifications}
\vspace{-1mm}
\label{sec:merge_spec}

The merge, also known as strong merge, operator of two contracts \(\mathcal{C}_1\) and \(\mathcal{C}_2\) is defined as follows,
\vspace{-2mm}
\begin{equation}
   \mathcal{C}_1 \cdot \mathcal{C}_2 = (a_1 \land a_2, (a_1 \land a_2)\rightarrow [(a_1\rightarrow g_1) \land (a_2 \rightarrow g_2)])
   \vspace{-1mm}
\end{equation}
In addition to \emph{strong merge}, contract theory defines other operators over a pair of contracts such as \emph{composition} and \emph{conjunction}~\cite{benveniste2018contracts,passerone2019coherent}. Among all these operators, strong merge is the only operator that requires assumptions from both unit contracts (and as a result, unit test specifications) to hold true. Thus, we choose the strong merge operator to derive the merged test specification.
Given two unit test specifications, \(\varphi_{\text{test},1}\) and \(\varphi_{\text{test},2}\), we can construct the corresponding contracts \(\mathcal{C}_1 = (a_1, a_1 \rightarrow g_1)\) and \(\mathcal{C}_2 = (a_2, a_2 \rightarrow g_2)\), where \(a_i = (\varphi^{init}_\text{sys} \wedge \square \varphi^{s}_\text{sys} \wedge \square \lozenge \varphi^{f}_\text{sys})\) being the assumptions on the system (under test), and \(g_i = (\varphi^{init}_{\text{test}, i} \wedge \square \varphi^{s}_{\text{test}, i} \wedge \square \lozenge \varphi^{f}_{\text{test}, i} \wedge \square \psi^{s}_{\text{test}, i}\wedge \square \lozenge \psi^{f}_{\text{test}, i})\) being the guarantees for unit test \(i\). 
\begin{remark}
We make the following modifications to guarantees \(g_i\) for brevity. First, we assume that the only recurrence requirements in the test specification is \(\square \lozenge \psi^{f}_{\text{test},i}\), which is not a part of the system's assumptions on the environment. Second, we assume that the merged test environment  \(\mc{T}_{\text{test},m}\) is a simple product transition system of the unit test environments, \(\mc{T}_{\text{test},1}\) and \(\mc{T}_{\text{test},2}\). On the merged test environment, we assume that the initial conditions  \(\varphi^{init}_{\text{test}, 1}\) and  \(\varphi^{init}_{\text{test}, 2}\) are equivalent, and test environment dynamics \(\varphi^{s}_{\text{test}, 1}\) and \(\varphi^{s}_{\text{test}, 2}\) are equivalent. Therefore, in merging the two unit specifications, we refer to the test guarantees as \(g_{t,i} = \square \psi^{s}_{\text{test}, i}\wedge \square \lozenge \psi^{f}_{\text{test}, i}\).
\end{remark}
\begin{definition2}[Merged Test]
From the merged contract \(\mc{C}_m := (a_m, a_m \rightarrow g_m) = \mc{C}_1 \cdot \, \mc{C}_2\), 
the specification \(\varphi_{\text{test},m} = a_m \rightarrow g_m\), where \(a_m = a_1 \land a_2\), and \(g_m = [(a_1\rightarrow g_1) \land (a_2 \rightarrow g_2)]\) is the \textit{merged test specification}. A test environment policy \(\pi_{\text{test}, m}\) for merged test specification \(\varphi_{\text{test},m}\) results in a test execution \(\sigma \models \varphi_{\text{test},m}\).
\end{definition2}
\vspace{-1mm}
\begin{lemma}
\label{lm:strong_merge}
Given unit test specifications \(\varphi_{\text{test},1}\) and \(\varphi_{\text{test},2}\) such that \(\varphi_{\text{test},m} = a_m \rightarrow g_m\) is the corresponding merged test specification. Then, for every test execution \(\sigma \models \varphi_{\text{test},m}\) such that \(\sigma \models a_m\), we also have that \(\sigma \models \varphi_{\text{test},1}\) and \(\sigma \models \varphi_{\text{test},2}\).
\vspace{-1mm}
\end{lemma}
\begin{proof}
Suppose \(C_1\) and \(C_2\) are the assume-guarantee contracts corresponding to unit test specifications \(\varphi_{\text{test},1}\) and \(\varphi_{\text{test},2}\). Applying strong merge operator on contracts \(C_1\) and \(C_2\), we get:
\vspace{-1mm}
\begin{equation}
\label{eq:general_test_spec}
\begin{split}
    \mathcal{C}_1 \cdot \mathcal{C}_2 = & (a_1 \land a_2, (a_1 \land a_2)\rightarrow [(a_1\rightarrow g_1) \land (a_2 \rightarrow g_2)]) \\ = & \big(a_1 \land a_2, \neg a_1 \lor \neg a_2 \lor (g_1 \land g_2)\big) .
\end{split}
\end{equation}
Thus, the merged test specification \(\varphi_{\text{test},m} = \neg a_1 \lor \neg a_2 \lor (g_1 \land g_2)\) requires either one of the assumptions to not be satisfied, or for both the guarantees hold. Since \(\sigma \models a_m = a_1 \land a_2\), and \(\sigma \models \varphi_{\text{test},m}\), we get that \(\sigma \models \varphi_{\text{test},1}\) and \(\sigma \models \varphi_{\text{test},2}\).
\qed
\end{proof}

A key point in our framework is that we select $g_1$ and $g_2$ to guide the test search, that is, we do not allow merged test policies that vacuously satisfy the merged test specification. This allows the test environment to always give the system an opportunity to satisfy its specification. If assumptions ever get violated, that is because of the system, and not the design of the test.

Returning to our lane change example, we define the unit test specifications as merging \textit{behind} a car and merging \textit{in front} of a car. The respective saturated assume guarantee contracts are defined as $\mathcal{C}_1 = (a_1, a_1 \rightarrow g_1)$ and $\mathcal{C}_2 = (a_2, a_2 \rightarrow g_2)$ with $a_1=\varphi^{init}_{sys} \land \square \varphi^{s}_{sys} \land \square \lozenge (y=2)$ and $g_1=\square \lozenge (y=y_1=2\land x=x_1+1)$, and $a_2=\varphi^{init}_{sys} \land \square \varphi^{s}_{sys} \land \square \lozenge (y=2)$ and $g_2=\square \lozenge (y=y_2=2\land x=x_2-1)$ being the assumptions and guarantees of the two individual tests.
Thus, after applying the strong merge operation to the two contracts, the guarantee of the merged test specification for the lane change example is,
\vspace{-2mm}
\begin{equation}
\label{eq:lanechangespec}
    g_m = \square \lozenge (y=y_1=2\land x=x_1+1) \land \square \lozenge (y=y_2=2\land x=x_2-1) .
    \vspace{-2mm}
\end{equation}
\subsection{Temporal Constraints on the Merged Test Specification}
\label{sec:temporal_const}
\begin{definition2}[Temporally constrained tests]
For a test trace $\sigma$, let $\sigma_t$ be the suffix of the trace, starting at time $t$. Let $t_{S1},t_{S2}$ be times such that $\sigma_{t_{S1}}\models \varphi_{\text{test},1}$ and $\sigma_{t_{S2}}\models \varphi_{\text{test},2}$, and assume there exists a time $t_{F1}$ such that $t_{F1}=min(t)$ for all $t$, $t>t_{S1}$ such that $\sigma_{t_F1}\not\models \varphi_{\text{test},1}$ and assume that there exists a time $t_{F2}$ such that $t_{F2}=min(t)$ for all $t$, $t > t_{S2}$ such that $\sigma_{t_{F2}}\not\models \varphi_{\text{test},2}$. Then if $t_{S1}=t_{S2}=t_1$ and $t_{F1}=t_{F2}=t_2$ the tests are \textit{parallel-merged} in the interval $t \in [t_1,t_2]$. If $t_{S1}<t_{S2}$ and $t_{F1}<t_{F2}$, or $t_{S1}>t_{S2}$ and $t_{F1}>t_{F2}$, the tests are \textit{temporally constrained}.
\end{definition2}
In this section, we will outline when the merged test specification requires a more constrained temporal structure.
To ensure that the test execution will provide the desired information, we need to make certain that each test specification is sufficiently checked.
For example, consider the lane change example. There exist many executions in which one of the unit tests is satisfied (i.e. the car merges in front of a vehicle), but it is not guaranteed that the other specification is satisfied as well. Therefore these two tests can be parallel-merged. In contrast to this there exist test specifications where satisfying one will trivially satisfy the other. Then we are not able to distinguish which specification was checked, thus these unit tests should not be parallel-merged to ensure that during the test there is a point in time where each test specification is satisfied individually.
\begin{proposition}
\label{prop:temp_con}
If for two test specifications $\varphi_{\text{test},1}$ and $\varphi_{\text{test},2}$, and the set of all test executions $\Sigma$, we have $\sigma\models \varphi_{\text{test},1} \iff \sigma\models \varphi_{\text{test},2} \: \forall \: \sigma \in \Sigma$, then these tests cannot be parallel-merged. Instead, the temporal constraint must be enforced on $g_{t,1}$ and $g_{t,2}$.
\end{proposition}
\begin{proof}
We refine the general specification in equation~(\ref{eq:general_test_spec}), which allows any temporal structure, to include the temporal constraints in the guarantees. The temporally constrained merged test specification is thus defined as $\varphi_{\text{test},m}' = a_m \rightarrow g_m'$, with
\vspace{-2mm}
\begin{equation}
\label{eq:refined_spec}
g_m' = (\neg a_1 \lor \neg a_2 \lor (\lozenge(g_{t,1} \land \neg g_{t,2})\land \lozenge(\neg g_{t,1} \land g_{t,2})\land (g_1 \land g_2))).
\end{equation}

Because any trace $\sigma$ satisfying $\varphi_{\text{test},m}'$ will also satisfy $\varphi_{\text{test},m}$, $\sigma \models \varphi_{\text{test},m}' \Rightarrow \sigma \models \varphi_{\text{test},m}$.
Any test trace satisfying this specification will consist of at least one occurrence of visiting a state satisfying $g_{t,1}$ and not $g_{t,2}$ and vice versa. Thus the guarantees of the specifications for each unit test, $g_{t,1}$ and $g_{t,2}$ are checked individually during the merged test which satisfies the temporal constraints.\qed
\end{proof}
\subsection{Receding Horizon Synthesis of Test Policy Filter}
Since the test specification characterizes the set of possible test executions, we need a policy for the test environment that is consistent with the test specification. In this section, we detail the construction of an auxiliary game graph and algorithms for receding horizon synthesis of the test specification on the auxiliary game graph. This filter will then be used to find the test policy (detailed in Section~\ref{sec:mcts}). 
\label{sec:rhc}
\vspace{-5mm}
\subsubsection{Auxiliary Game Graph \(G_\text{aux}\)}
\label{sec:graph_construction}
\begin{figure}
  \begin{minipage}[c]{0.55\textwidth}
  \vspace{-2mm}
    \includegraphics[width=0.95\textwidth]{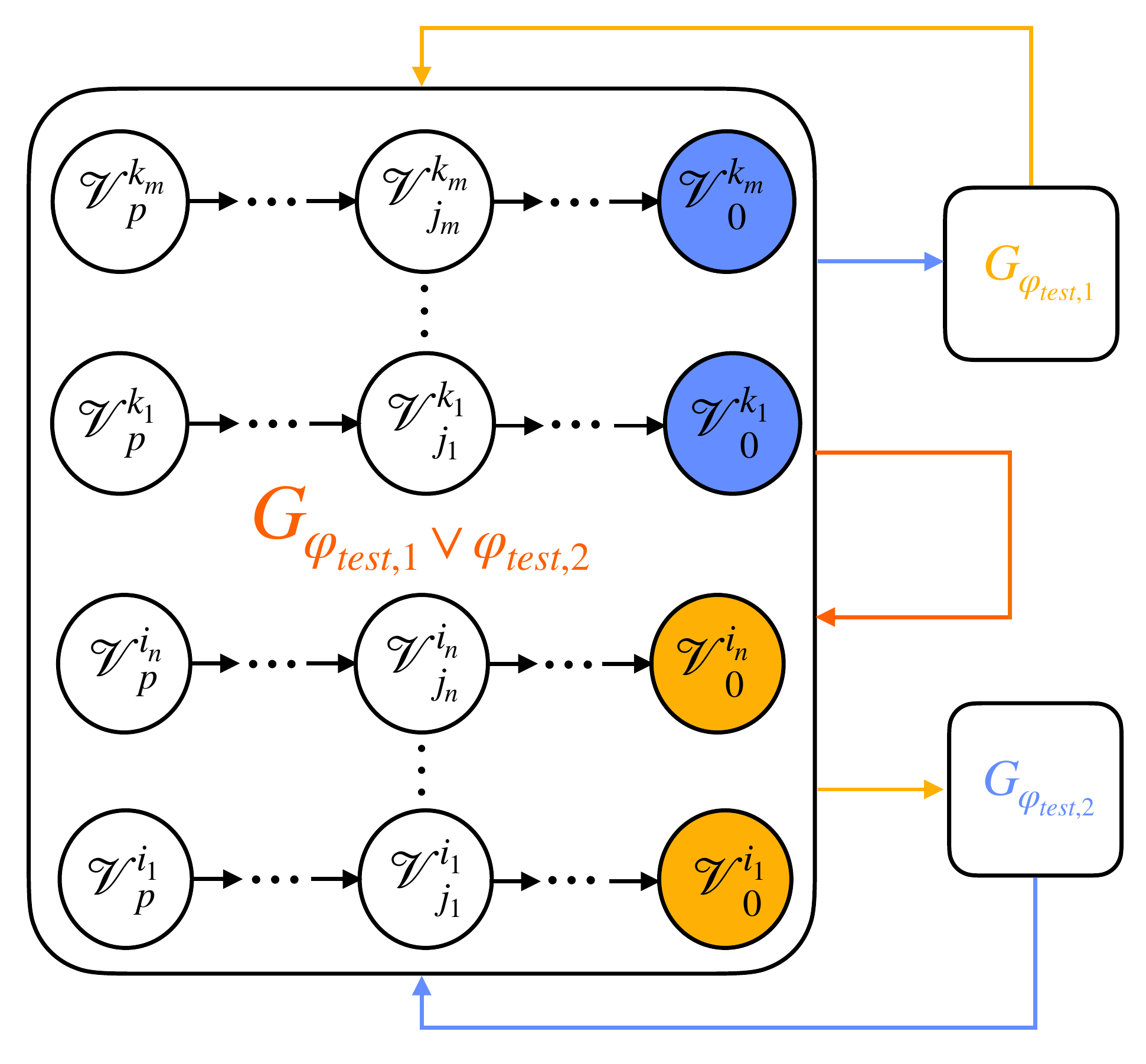}
  \end{minipage}\hfill
  \begin{minipage}[c]{0.45\textwidth}
    \caption{Auxiliary game graph construction for the merged test specification of unit test specifications \(\varphi_{\text{test},1}\) and \(\varphi_{\text{test},2}\). Subgraphs \(G_{\varphi_{\text{test},1} \vee \varphi_{\text{test},2}}\), \(G_{\varphi_{\text{test},1}}\) and \(G_{\varphi_{\text{test},2}}\) are copies of the game graph \(G\) constructed per Definition~\ref{def:game_graph}. In \(G_{\varphi_{\text{test},1} \vee \varphi_{\text{test},2}}\), the sets of states at which the progress propositional formulas of test specifications, \(\varphi_{\text{test},1}\) and \(\varphi_{\text{test},2}\), are satisfied are shaded yellow and blue, respectively.
    } \label{fig:graph}
  \end{minipage}
\vspace{-8mm}
\end{figure}

Assume we are given a game graph \(G = (V,E)\) constructed according to Definition~\eqref{def:game_graph}, and a (merged) test specification \(\varphi_{\text{test},m}\) in $GR(1)$ form as in equation~\eqref{eq:test_spec}. Then, for each recurrence requirement in the test specification, \(\square \lozenge \psi^{f}_{\text{test}}\), we can find a set of states \(\mc{I} = \{i_1, \ldots, i_n\} \subseteq V\) that satisfy the propositional formula \(\psi^{f}_{\text{test}}\). For each \(i \in \mc{I}\), there exists a non-empty subset of vertices \(V^s \subseteq V\) that can be partitioned into \(\{\mc{V}^i_0, \ldots, \mc{V}^i_n\}\). We follow~\cite{wongpiromsarn2012receding} in partitioning the states; \(\mc{V}^i_k\) is the set of states in \(V\) that is exactly \(k\) steps away from the goal state \(i\). From this partition of states, we can construct a partial order, \(\mc{P}^i = (\{\mc{V}^i_0, \ldots, \mc{V}^i_n\}, \leq)\), such that \(\mc{V}^i_l \leq \mc{V}^i_{l-1}\) for all \(l \in \{0,\ldots,n\}\). This partial order will be useful in the receding horizon synthesis of the test policy outlined below~\cite{wongpiromsarn2012receding}. 
We construct an auxiliary game graph \(G_\text{aux} = (V_\text{aux}, E_\text{aux})\) (illustrated in Figure~\ref{fig:graph}) to accommodate any temporal constraints on the merged test specification before proceeding to synthesize a filter for the test policy. Without loss of generality, we elaborate on the auxiliary graph construction in the case of one recurrence requirement in each unit specification, but this approach can be easily extended to multiple progress requirements. An illustration of the auxiliary graph is given in Figure~\ref{fig:graph}. Let \(\varphi_{\text{test},1}\) and \(\varphi_{\text{test},2}\) be the two unit test specifications, with \(\psi^f_{\text{test},1}\) and \(\varphi^f_{\text{test},2}\), respectively. First, we make three copies of the game graph \(G = (V,E)\) --- \(G_{\varphi_{\text{test},1} \vee \varphi_{\text{test},2}} = (V_{1 \vee 2}, E_{1 \vee 2})\), \(G_{\varphi_{\text{test},1}} = (V_{1}, E_{1})\), and \(G_{\varphi_{\text{test},2}} = (V_{2}, E_{2})\). Note that, \(V_{1 \vee 2}\), \(V_{1}\) and \(V_{2}\) are all copies of \(V\), but are denoted differently for differentiating between the vertices that constitute \(G_{\text{aux}}\), and a similar argument applies to edges of these subgraphs. Let \(\mc{V}^i_0 = \bigcup \mc{V}^{i_j}_0 \subseteq V_{1 \vee 2}\) be the set of states in \(G_{\varphi_{\text{test},1} \vee \varphi_{\text{test},2}}\) that satisfy propositional formula \(\psi^f_{\text{test},1}\). Likewise, the set of states \(\mc{V}^k_0 \subseteq V_{1 \vee 2}\) satisfy the propositional formula \(\psi^f_{\text{test},2}\). 
\vspace{-5mm}

Now, we connect the various subgraphs through the vertices in \(\mc{V}^i_0\) and \(\mc{V}^k_0\). Let \((v^k_0, u)\) be an outgoing edge from a node \(v^k_0 \in \mc{V}^k_0\), and let \(u_1\) be the vertex in subgraph \(G_{\text{test},1}\) that corresponds to vertex \(u\) in \(G_{\varphi_{\text{test},1} \vee \varphi_{\text{test},2}}\). Remove edge \((v^k_0, u)\) and add the edge \((v^k_0, u_1)\). Likewise, every outgoing edge from \(\mc{V}^i_0 \cup \mc{V}^k_0\) in \(G_{\varphi_{\text{test},1} \vee \varphi_{\text{test},2}}\) is replaced by adding edges to \(G_{\varphi_{\text{test},1}}\) and \(G_{\varphi_{\text{test},2}}\). On subgraphs \(G_{\varphi_{\text{test},1}}\) and \(G_{\varphi_{\text{test},2}}\), vertices are partitioned and partial orders are constructed once again for \(\psi^f_{\text{test},1}\) and \(\psi^f_{\text{test},2}\), respectively. From \(\mc{V}^i_0\) defined on the nodes of the graph \(G_{\varphi_{\text{test},1}}\), every outgoing edge is replaced by a corresponding edge to \(G_{\varphi_{\text{test},1} \vee \varphi_{\text{test},2}}\). Subgraph \(G_{\varphi_{\text{test},2}}\) is connected back to \(G_{\varphi_{\text{test},1} \vee \varphi_{\text{test},2}}\) in a similar manner. The construction of the auxiliary graph \(G_\text{aux}\) and partial order \(\mc{P}^i\) is summarized in Algorithm~\ref{alg:partial_order}. Our choice of constructing the auxiliary graph in this manner is amenable to constructing a simple partial order as outlined below.
\begin{assumption}
 \label{assumption}
 For unit test specifications \(\varphi_{\text{test},1}\) and \(\varphi_{\text{test},2}\) with recurrence specifications \(\varphi^p_1\) and \(\varphi^p_2\), respectively, such that \(\varphi^p_1 = \square \lozenge \psi^f_{\text{test},1}\) and \(\varphi^p_2 = \square \lozenge \psi^f_{\text{test},2}\). Suppose there exist partial orders \(\mc{P}^i = (\{\mc{V}^i_n, \ldots, \mc{V}^i_0\}, \leq)\) and \(\mc{P}^k = (\{\mc{V}^k_m, \ldots, \mc{V}^k_0\}, \leq)\) on \(G\) corresponding to \(\psi^f_{\text{test},1}\) and \(\psi^f_{\text{test},2}\), respectively. Assume that at least one of the following is true: (a) there exists an edge \((u_1,v_{2})\) where \(u_1 \in \mc{V}^i_0\) and \(v_{2} \in \mc{V}^k_j\) for some \(j \in 1,\ldots, m\), (b) there exists an edge \((u_2,v_{1})\) where \(u_2 \in \mc{V}^k_0\) and \(v_{1} \in \mc{V}^i_j\) for some \(j \in 1,\ldots, n\). 
\end{assumption}
 \begin{lemma}
\label{lm:test_spec}
If Assumption~\ref{assumption} holds, there exists a partial order on \(G_\text{aux}\) for the merged recurrence propositional formula, \(\psi^f_{\text{test},m}\), where \(\psi^f_{\text{test},m}\) is the propositional formula that evaluates to true at: (i) all \(v \in V_{1 \vee 2}\) such that \(v \models \psi^f_{\text{test},1} \wedge \psi^f_{\text{test},2}\), (ii) all \(v \in V_{1}\) such that \(v \models \psi^f_{\text{test},1}\), and (iii) all \(v \in V_{2}\) such that \(v \models \psi^f_{\text{test},2}\).
\end{lemma}
\begin{proof}
Let \(\mc{V}^m_0 \subseteq V_{\text{aux}}\) denote the non-empty set of states at which \(\psi^f_{\text{test},m}\) evaluates to true. Then, let \(\mc{V}^m_j \subseteq V_{\text{aux}}\) be the subset of states that is at least \(j\) steps away from a vertex in \(\mc{V}^m_0\). Then, we can construct the partial order \(\mc{P}^m = (\{\mc{V}^m_l, \ldots, \mc{V}^m_0\}, \leq)\), where \(l\) is the distance of the farthest vertex connected to \(\mc{V}^m_0\). The subset of vertices \(\bigcup_j \mc{V}^m_j \subseteq V_\text{aux}\) is non-empty because \(\mc{V}^m_0\) is non-empty. Furthermore, from Assumption~\ref{assumption}, if (a) holds, there exists a \(j \in \{1, \ldots, l\}\) such that \(\mc{V}^m_j \cap \mc{V}^i_0\) is non-empty. Likewise, if (b) holds, there exists a \(j \in \{1, \ldots, l\}\) such that \(\mc{V}^m_j \cap \mc{V}^k_0\) is non-empty. Therefore, for some \(j\in \{1, \ldots, l\}\) there exists a test execution \(\sigma\) over the game graph \(G_{\text{aux}}\) such that \(\sigma \models \square \lozenge \psi^f_{\text{test},m}\).
\qed
\end{proof}
\begin{remark}
If Assumption~\ref{assumption} is not true, the unit tests corresponding to test specifications \(\varphi_{\text{test},1}\) and \(\varphi_{\text{test},2}\) cannot be merged. 
\end{remark}
\subsubsection{Receding Horizon Synthesis on \(G_\text{aux}\)}
 We leverage receding horizon synthesis to scalably compute the set of states \(\mc{W}\) from which the test environment can realize the test specification on the system in a test execution. Note that we are not synthesizing a test strategy using the receding horizon approach, instead using \(\mc{W}\) as a filter on a search algorithm (MCTS) that finds an optimal test policy. Further details on applying receding horizon strategies for temporal logic planning can be found in~\cite{wongpiromsarn2012receding}. A distinction in our work is that there can be multiple states in graph \(G_\text{aux}\) that satisfy a progress requirement on the test specification. 
 
 For a test specification \(\varphi_{\text{test},1}\) with progress propositional formula \(\square \lozenge \psi^f_{\text{test},1}\), let \(\mc{I}\) be the set of states on \(G_\text{aux}\) at which \(\psi^f_{\text{test},1}\) evaluates to true.
 Specifically, for some goal \(i \in \mc{I}\), if the product state starts at \(j\) steps from \(i\) (i.e. \(v\in \mc{V}^i_{j+1}\)), the test environment is required to guide the product state to \(\mc{V}^i_{j-1}\). The corresponding formal specification for the test environment is,
 \vspace{-2mm}
\begin{equation}
\psi^i_j = (v \in \mathcal{V}^i_{j+1} \wedge \Phi \wedge \square \varphi^s_{\text{sys}} \wedge \square \lozenge \varphi^f_{\text{sys}}) \rightarrow (\square \lozenge (v \in \mathcal{V}^i_{j-1}) \wedge \square \varphi^s_{\text{test}} \wedge \square \psi^s_{\text{test}} \wedge \square \Phi),
\vspace{-2mm}
\end{equation}
where \(\Phi\) is the invariant condition that ensures that \(\psi^i_j\) is realizable. See~\cite{wongpiromsarn2012receding} for further details on how this invariant can be constructed. 
Since there are \(|\mc{I}|\) different ways to satisfy the goal requirement \(\psi^f_{\text{test},1}\), and the test specification requires that we satisfy \(\psi^f_{\text{test},1}\) for at least one \(i \in \mc{I}\). To capture this in the receding horizon framework the test execution must progress to at least one \(i\in \mc{I}\), formally stated as,
\vspace{-2mm}
\begin{equation}
\label{eq:disjunct}
\Psi^{\mc{I}}_j = \vee_{i\in \mathcal{I}} \, \psi^i_j\, .
\vspace{-1mm}
\end{equation}
Thus, the set of states from which the test environment has a strategy that satisfies the specification in equation~\eqref{eq:disjunct} is the short horizon filter, denoted by \(\mc{W}^{\mc{I}}_j\). Let \(j_\text{max}\) denote the supremum of all shortest paths from a vertex \(v\in V\) to some \(i \in \mc{I}\). Then, overall test policy filter is the union of short-horizon test policy filters,
\vspace{-2mm}
\begin{equation}
    \label{eq:winning_set}
    \mc{W}^{\mc{I}} = \bigcup_{j=1}^{j_\text{max}} \mc{W}^{\mc{I}}_j \,.
    \vspace{-2mm}
\end{equation}
The synthesis of \(\mc{W}^{\mc{I}}\) and its use as a test policy filter in the MCTS procedure used to find the test environment policy is outlined in Algorithm~\ref{alg:algorithm}. Note, that this receding horizon approach to generating a filter \(\mc{W}\) can be applied on any $GR(1)$ specification and its corresponding game graph. For the merged test specification, \(\mc{W}^{\mc{I}}\) is generated on \(G_\text{aux}\) where \(\mc{I}\) is the set of states corresponding to \(\psi^f_{\text{test},m}\), and for simplicity, we apply the following arguments on \(G_\text{aux}\). Let \(G_{\mc{W}^{\mc{I}}} = (V_\mc{W}, E_\mc{W})\) be the subgraph of \(G_\text{aux}\) induced by \(\mc{W}^{\mc{I}}\) such that \(V_\mc{W} = \mc{W}^{\mc{I}} \subseteq V_\text{aux}\) and \(E_\mc{W} = \{(u,v) \in E_\text{aux}|\, u\in \mc{W}^{\mc{I}} \wedge v\in \mc{W}^{\mc{I}}\}\). 
\vspace{-3mm}
\subsubsection{On \(\mc{W}^{\mc{I}}\) as a test policy filter} Inspired by work on shield synthesis~\cite{bloem2015shield}, we use the winning set \(\mc{W}^{\mc{I}}\) as a filter to guide rollouts in the Monte Carlo Tree Search sub-routine for finding the test policy. Since \(\Psi^{\mc{I}}_j\) is a disjunction of short-horizon $GR(1)$ specifications, it is possible that an execution always satisfies \(\Psi^{\mc{I}}_j\) without ever satisfying the progress requirement \(\square \lozenge \psi^f_{\text{test}}\). This happens when the test execution makes progress towards some \(i\in \mc{I}\) but never actually reaches a goal in \(\mc{I}\), resulting in a live lock. Further details addressing this are given in the Appendix. We \textit{assume} that the graph is constructed such that there are no such cycles. In addition to using $W^\mathcal{I}$ to ensure that \(\Psi^{\mc{I}}_j\) will always be satisfied, we enforce progress by only allowing the search procedure to take actions that will lead to a state which is closer to one of the goals $i \in \mathcal{I}$. Thus, the search procedure will ensure that for every state $v_l \in \mathcal{V}_j^i$, the control strategy for the next horizon will end in $v_{l'} \in \mathcal{V}_{k}^i$, such that $k\leq l$ for at least one goal $i \in \mathcal{I}$.
\vspace{-3mm}
\begin{figure}
\begin{minipage}[c]{0.65\textwidth}
\centering
\vspace{-1mm}
\caption{Illustration of the intersection of the winning sets for the unit specification. $V_{\text{test}}$ are depicted as circles and $V_{sys}$ as rhombi. The black states lie in the intersection and the filter will ensure that only these states are being searched. The orange intersection represents the set of traces of the merged test specification.}
\label{fig:intersection_ws}
\end{minipage}\hfill
\begin{minipage}[c]{0.3\textwidth}
\vspace{-5mm}
\includegraphics[width=0.9\textwidth]{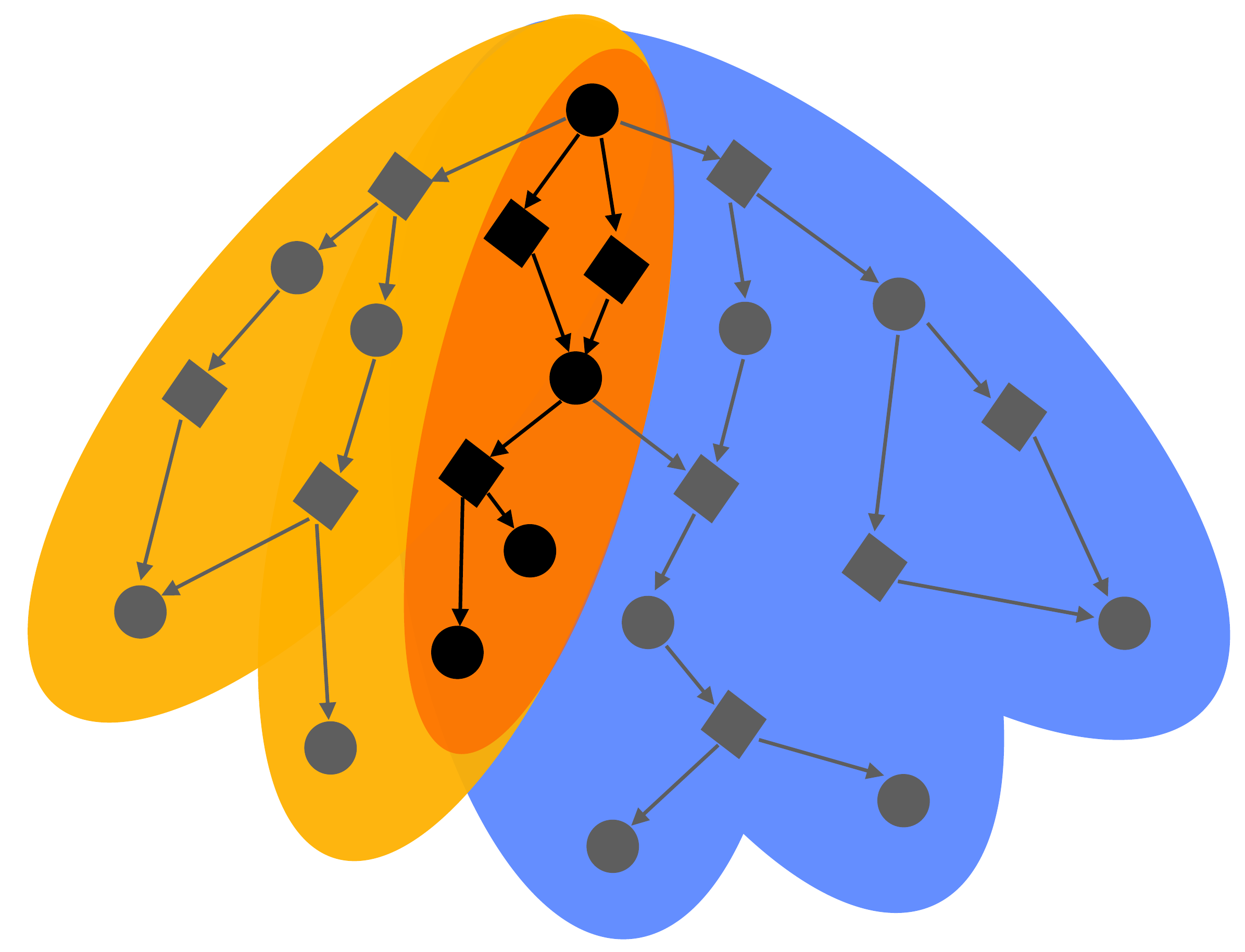}
\end{minipage}
\vspace{-9mm}
\end{figure}
\begin{theorem}
Receding horizon synthesis of test filter \(\mc{W}^{\mc{I}}\) is such that any test execution \(\sigma\) on \(G_{\mc{W}^{\mc{I}}}\) starting from an initial state in $V_\mathcal{W}\cap V$ satisfies the test specification in equation~\eqref{eq:test_spec}.
\label{thm:thm1}
\end{theorem}
\begin{proof}
For the recurrence formula of the merged test specification, \(\square \lozenge \psi^f_{\text{test}, m}\), suppose there exists a single vertex on \(G_\text{aux}\) that satisfies \(\psi^f_{\text{test},m}\). Then, it is shown in~\cite{wongpiromsarn2012receding} that if there exists a partial order \((\{\mc{V}^i_p, \ldots, \mc{V}^i_0\}, \leq)\) on $G_\text{aux}$, we can find a set of vertices \(\mc{W}^{i} \subseteq V_\text{aux}\), such that every test execution $\sigma$ that remains in \(\mc{W}^{i}\), will satisfy the safety requirements $\square \varphi^s_{\text{test}}$ and $\square \psi^s_{\text{test}}$, and the invariant $\Phi$. Furthermore, given the partial order \((\{\mc{V}^i_p, \ldots, \mc{V}^i_0\}, \leq)\), one can find a test policy to ensure that the \(\sigma\) makes progress along the partial order such that for some \(t >0\), \(\sigma_t \in \mc{V}^i_0\). However, in case of multiple vertices in \(G_\text{aux}\) that satisfy \(\psi^f_{\text{test},m}\), we need to extend the receding horizon synthesis to specification \(\Psi^{\mc{I}}_j\). We construct the filter \(\mc{W}^{\mc{I}}\) and also check that for every test execution $\sigma$, there exists $i \in \mathcal{I}$ such that for every $k\geq 0$, \(\sigma_k \in \mc{V}^i_j\) and \(\sigma_{k+1} \in \mc{V}^i_{j'}\). Therefore, because the auxiliary game graph is assumed to not have cycles, the test execution makes progress on the partial order of at least one \(i \in \mc{I}\) at each timestep, thus eventually satisfying $\psi^f_{\text{test},m}$. Thus every execution of our algorithm will satisfy equation~(\ref{eq:test_spec}).\qed
\end{proof}
\vspace{-8mm}
\begin{algorithm}
 \caption{Merge Unit Tests $(\varphi_{\text{test},1}, \varphi_{\text{test},2}, \varphi_{\text{sys}}, \mc{T}_\text{sys}, \mc{T}_{\text{test},1}, \mc{T}_{\text{test},2}, \rho)$}
 \begin{algorithmic}[1]
 \renewcommand{\algorithmicrequire}{\textbf{Input:}}
 \renewcommand{\algorithmicensure}{\textbf{Output:}}
 \REQUIRE Unit test specifications \(\varphi_{\text{test},1}\) and \(\varphi_{\text{test},2}\), system specification  \(\varphi_{\text{sys}}\), System \(\mc{T}_\text{sys}\), unit test environments, \(\mc{T}_{\text{test},1}\) and  \(\mc{T}_{\text{test},2}\), and quantitative metric of robustness \(\rho\), 
 \ENSURE   Merged test specification $\varphi_{\text{test}, m}$, Merged test environment \(\mc{T}_{\text{test}, m}\), Merged test policy \(\pi_{\text{test}, m}\) 
  \STATE $\mathcal{C}_1,\, \mathcal{C}_2 \leftarrow $ Construct contracts for $\varphi_{\text{test},1}$ and \(\varphi_{\text{test},2}\)
   \STATE \(\mc{T}_\text{test} \leftarrow \mc{T}_{\text{test},1} \times \mc{T}_{\text{test},2}\) Merged test environment 
   \STATE \(\mc{T}_\text{prod} \leftarrow \mc{T}_\text{sys} \times \mc{T}_\text{test}\) Product transition system 
  \STATE \(G \leftarrow\) Game graph from product transition system \(\mc{T}_{prod}\)
  \STATE \(\mathcal{C}_m := (a_m, a_m \rightarrow g_m) \leftarrow \) strong merge\((\mathcal{C}_1, \,\mathcal{C}_2\)) Constructing the merged specification
  \STATE \(\varphi_{\text{test},m} \leftarrow a_m \rightarrow g_m\) Merged test specification
  \STATE \(G_\text{aux} \leftarrow\) Auxiliary game graph.
  \STATE \(\mathcal{I} = \{s \in \mc{V}_{\text{aux}}| s \models \psi^f_{\text{test},m}\}\) Defining goal states and partial orders
  \FOR {$i \in \mathcal{I}$}
  \STATE \(\mathcal{P}^i := \{(\mathcal{V}^i_p, \ldots, \mathcal{V}^i_0)\} \leftarrow\) Partial order for goal $i$
  \STATE \(\psi^i_j \leftarrow\) Receding horizon specification for goal \(i\) at distance \(j\)
  \ENDFOR
  \STATE \(\mc{W}^{\mc{I}} := \{\mathcal{W}^i_j\} \leftarrow\) Test policy filter for goal $i$ at a distance of $j$ 
  \STATE \(\pi_{\text{test},m} \leftarrow \) Searching for test policy guided by \(\mc{W}^{\mc{I}}\)
 \RETURN $\varphi_{\text{test}, m}$, \(\mc{T}_{\text{test}, m}\), \(\pi_{\text{test}, m}\)
 \end{algorithmic} 
 \label{alg:algorithm}
 \end{algorithm}
\setlength{\textfloatsep}{4pt}
\vspace{-5mm}
\subsection{Searching for a Test Policy}


\label{sec:mcts}
To find the merged test policy $\pi_{text,m}$, we use Monte-Carlo Tree Search (MCTS), which is a search method that and combines random sampling with the precision of a tree search. 
Using MCTS with an upper confidence bound (UCB) was introduced in~\cite{kocsis2006bandit} as upper confidence bound for trees (UCT) which guarantees that given enough time and memory, the result converges to the optimal solution.
We use MCTS to find $\pi_\text{test,m}^*$, the approximate solution to Problem~\ref{problem_statement} for the merged test.
We apply the filter that was generated according to the approach detailed in Section~\ref{sec:rhc} to constrain the search space as shown graphically in Figure~\ref{fig:intersection_ws}.
\begin{proposition} Algorithm 1 is sound. 
\end{proposition}
\begin{proof}
This follows by construction of the algorithm and the use of MCTS with UCB.
Given a test policy \(\pi_\text{test}\) and a system policy \(\pi_\text{sys}\), for every resulting execution \(\sigma_{\pi_\text{sys} \times \pi_\text{test}}\) starting from an initial state in \(\mc{W}^{\mc{I}}\), it is guaranteed that $\sigma \models \varphi_{\text{test},m}$ by Theorem~\ref{thm:thm1}. This is because for any action chosen by the test environment according to the policy \(\pi_\text{test}\) found by MCTS, we are guaranteed to remain in \(\mathcal{W}^\mathcal{I}\) for any valid system policy \(\pi_\text{sys}\). If \(\mathcal{W}^\mathcal{I} = \emptyset\) or the initial state is not in \(\mathcal{W}^\mathcal{I}\), the algorithm will terminate before any rollout is attempted and no policy is returned. It can be shown that the probability of selecting the optimal action converges to 1 as the limit of the number of rollouts is taken to infinity. For convergence analysis of MCTS, please refer to~\cite{kocsis2006bandit}.
\qed
\end{proof}

\vspace{-6mm}
\subsubsection{Complexity Analysis}
\label{sec:complexity}
The time complexity of $GR(1)$ synthesis is in the order of $O(|N|^3)$, where $|N|$ is the size of the state space. To improve the scalability, our algorithm uses a receding horizon approach to synthesize the winning sets, which reduces the time complexity significantly, please prefer to~\cite{wongpiromsarn2012receding}.
The complexity for MCTS is given as $O(ijkl)$ with $j$ the number of rollouts, $k$ the branching factor of the tree, $l$ the depth of the tree, and $i$ the number of iterations. In our approach the filter reduces the size of the search space, for a visualization refer to Figure~\ref{fig:intersection_ws}. The number of rollouts and iterations are design variables, that can be chosen to ensure convergence. More details on the complexity of MCTS for the lane change example can be found in the Appendix.

\begin{definition2}[Coverage]
A test execution \(\sigma\) \emph{covers} a test specification \(\varphi_{\text{test}}\) if the test execution non-trivially satisfies the test specification, that is, \(\sigma \models \varphi_{\text{test}}\) and \(\sigma \models \varphi^{init}_\text{sys} \wedge \square \varphi^{s}_\text{sys} \wedge \square \lozenge \varphi^{f}_\text{sys}\). A set of test executions \(T = \{\sigma_1, \ldots, \sigma_n\}\) \emph{covers} the set of test specifications \(\Phi := \{\varphi_{\text{test},1}, \ldots, \varphi_{\text{test},m}\}\) iff for each test specification \(\varphi_{\text{test}} \in \Phi\), there exists a test execution \(\sigma_j \in T\) such that \(\sigma_j\) \emph{covers} \(\varphi_{\text{test},1}\).
\end{definition2}
Optimizing for the smallest set of test executions that cover a set of test specifications is combinatorial in the number of test specifications. In this work, we outlined an algorithm for merging two unit tests. In future work, given \(N\) unit tests, we will consider the problem of constructing a smaller set of \(N'\) merged test specifications with upper bounds on \(N'/N\).  
\begin{lemma}
\label{lemma:coverage}
Given a set of unit test specifications, \(\Phi_T := \{\varphi_{\text{test},1}, \ldots, \varphi_{\text{test},N}\}\) such that \(N\) test executions are are required to cover \(\Phi\), i.e. one test execution for each test specification, merging unit tests results in \(N'\) test executions that cover \(\Phi\) where \(N' \leq N\). The equality holds iff no two unit tests in \(\Phi\) can be merged. \end{lemma}
\begin{proof}
If at least a pair of test specifications in \(\Phi\) can be merged, it is possible to characterize a set of test specifications \(\Phi'\) such that the cardinality of \(\Phi'\), \(N'\), is always smaller than \(N\). If each test specification in \(\Phi'\) has a test execution, then we have \(N' < N\) test executions.
\qed
\end{proof}



\vspace{-3mm}
\section{Examples}
\label{sec:results}
We implemented the examples as a discrete gridworld simulation in Python, where the system controller is non-deterministic and the test agents follow the test policy generated by our framework. We use the Temporal Logic and Planning Toolbox (TuLiP) to synthesize the winning sets~\cite{wongpiromsarn2011tulip} and online MCTS to find the test policy. Videos of the results can be found in the linked GitHub repository.
\vspace{-3mm}
\subsection{Lane Change}
\label{ex:merge}
\begin{figure}
\includegraphics[width=\textwidth]{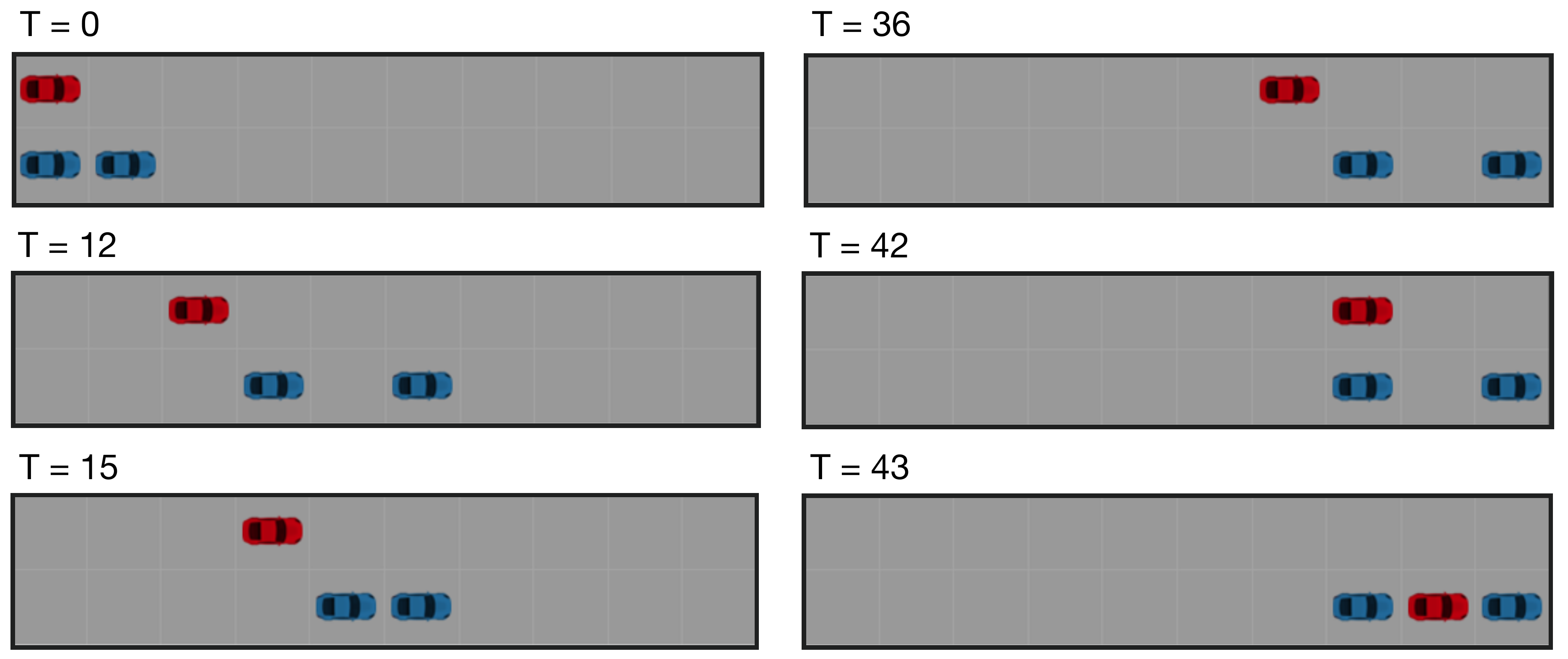}
\caption{Snapshots during the execution of the test generated by our framework. The system under test (red car) needs to merge onto the lower lane between the two test agents (blue cars).} \label{fig:merge_sim}
\end{figure}

For our discrete lane change example, we define $\rho(\sigma)$ as the x-value of the cell in which the system finished its lane change maneuver. We search for the test policy that satisfies the test specification in equation~(\ref{eq:lanechangespec}) as explained in Section~\ref{sec:approach}.
Snapshots of the resulting test execution are depicted in Figure~\ref{fig:merge_sim}.
\subsubsection{Unprotected left-turn at intersection}
\label{sec:intersection}
\begin{figure}
\begin{minipage}[c]{0.65\textwidth}
\centering
\includegraphics[width=0.95\textwidth]{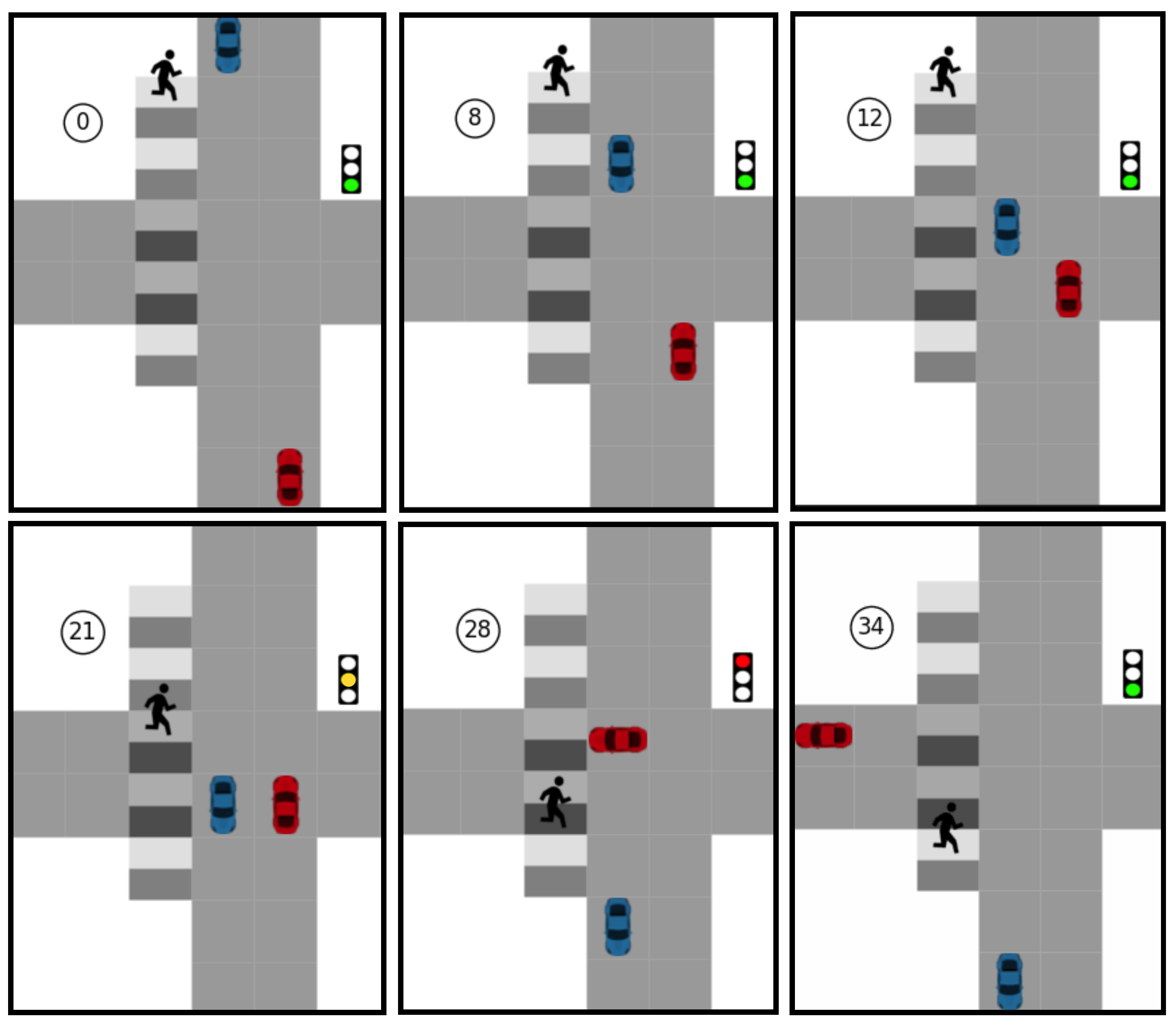}
\end{minipage}\hfill
\begin{minipage}[c]{0.35\textwidth}
\caption{Snapshots during the execution of the unprotected left turn test generated by our framework. The autonomous vehicle (AV) under test (red) should take an unprotected left turn and wait for the pedestrian and the car (blue) individually, which are agents of the test environment. In the snapshots at time steps 8 and 12, the AV waits just for the car, and in time step 21 it waits just for the pedestrian.} \label{fig:intersection_ex}
\end{minipage}
\vspace{5mm}
\end{figure}

Consider the example of an autonomous vehicle (AV) crossing an intersection with the intention of taking a left-turn. The test agents are a car approaching the intersection from the opposite direction and a pedestrian crossing the crosswalk to the left of the AV under test. The intersection layout can be seen in Figure~\ref{fig:intersection_ex}. The individual tests are defined to be waiting for a car, and waiting for a pedestrian while taking a left turn. The unit specification for waiting for the pedestrian are defined according to equation~(\ref{eq:test_spec}), with 
\vspace{-2mm}
\begin{equation}
\begin{split}
    \varphi^{init}_{sys} = (\textbf{x}_S \in I_S), \quad \square \lozenge \varphi_{sys}^f = \lozenge (\textbf{x}_S \in \mathcal{S}_G), \quad
     \square \lozenge \psi_{\text{test}}^f = \lozenge (\textbf{x}_S \in \mathcal{S}_P \land \textbf{x}_P \in T_P)\,,
\end{split}
\end{equation}
with $\textbf{x}_S$ the system coordinates, $I_S$ the initial state of the system, $\mathcal{S}_G$ the set of desired goal states after the left turn, $\textbf{x}_P$ the pedestrian coordinates, and $\mathcal{S_P}$ the states in which the car must wait for the pedestrian if the pedestrian is in a state in $T_P$. Similarly we define the specification for waiting for the tester car (detailed in the Appendix). 

The robustness metric is assumed to be the time until the traffic light changes to red starting the moment the system executes a successful left turn, and minimizing this metric results in a difficult test execution.
Next, we merge unit test contracts, and derive the resulting merged test specification. According to Proposition~\ref{prop:temp_con}, this merged specification needs to include the temporal constraints as defined in equation~(\ref{eq:refined_spec}).
In this example, waiting for the tester car and waiting for the pedestrian trivially imply each other in this example. Any execution of the system waiting at the intersection will satisfy both unit specifications. Thus we need to find a test where the system waits for just the tester car at some time during the test execution and waits for the tester pedestrian at another time during the test execution.
We follow the approach detailed in Section~\ref{sec:graph_construction} to generate the auxiliary graph for this example, with the terminal states corresponding to a successful left turn through the intersection after satisfying the temporally constrained merged test specification. The graph for this example is illustrated in Figure~\ref{fig:graph}, with $\text{test},1$ and $\text{test},2$ being the subscripts for the first and second unit test specification.
We then generate the test policy filter by constructing a partial order for the goal states and synthesizing the winning sets with the receding horizon strategy detailed in Section~\ref{sec:rhc}. Finally, applying this test filter on MCTS to find the test policy. Figure~\ref{fig:intersection_ex} shows snapshots from a test execution resulting from a test policy generated by Algorithm~\ref{alg:algorithm}. As expected, we see the system first waiting for the tester car to pass the intersection. Even after the tester car passes, the pedestrian is still traversing the crosswalk, causing the system to wait for the pedestrian, satisfying the temporally constrained merged test specification. 

\vspace{10mm}
\section{Conclusion and Future Work}
\label{sec:conclusion}
In this work, we presented a framework for merging unit test specifications. While we applied this framework to two discrete-state examples in the self-driving domain, this framework can be applied to test other autonomous systems as well. This paper details the mathematical and algorithmic foundation for merging two unit tests. This technique could be used as a subroutine to optimize for a small set of tests that cover several unit specifications. The winning set structure of the unit specifications could be leveraged to decide which unit specifications should be merged. The scalability of our algorithm can be further improved by symbolic implementations to synthesize the test policy filter. Lastly, we would like to show the results of this framework on continuous dynamical systems with a discrete abstraction for which the test policy filter can be synthesized. 

\paragraph{Acknowledgements}
We thank Dr. Ioannis Filippidis, Dr. Tichakorn Wongpiromsarn, {\'I}{\~n}igo {\'I}ncer Romeo, Dr. Qiming Zhao, Dr. Michel Ingham, and Dr. Karena Cai for valuable discussions that helped shape this work. The authors acknowledge funding from AFOSR Test and Evaluation program, grant FA9550-19-1-0302 and National Science Foundation award CNS-1932091.
\clearpage
%
%
%
\bibliographystyle{splncs04}
\bibliography{refs}
%
\section{Appendix}
\subsection{Construction of the Partial Order}
In Algorithm~\ref{alg:partial_order} we provide an algorithm to construct the partial order and the auxiliary game graph.
\begin{algorithm}
 \caption{Construction of Partial Order and Auxiliary Graph}
 \label{alg:partial_order}
 \begin{algorithmic}[1]
 \renewcommand{\algorithmicrequire}{\textbf{Input:}}
 \renewcommand{\algorithmicensure}{\textbf{Output:}}
 \REQUIRE Game graph $G = (V,E)$, propositional formulas \(\psi^f_{\text{test},1}\) and \(\psi^f_{\text{test},2}\) constituting the progress requirements of unit test specifications
 \ENSURE   Auxiliary game graph \(G_\text{aux}\)\\
  \STATE \(G_{\varphi_{\text{test},1} \vee \varphi_{\text{test},2}} := (V,E) \leftarrow G\) Initialize subgraph
  \STATE \(G_{\varphi_{\text{test},1}} := (V_{1},E_{1}) \leftarrow G\) Initialize subgraph
  \STATE \(G_{\varphi_{\text{test},2}} := (V_{2},E_{2}) \leftarrow G\) Initialize subgraph
  
  \STATE \([\mc{P}^i_{\varphi_{\text{test},1} \vee \varphi_{\text{test},2}}, \mc{P}^k_{\varphi_{\text{test},1} \vee \varphi_{\text{test},2}}] \leftarrow\) Partial order\((G_{\varphi_{\text{test},1} \vee \varphi_{\text{test},2}}, [\psi^f_{\text{test},1}, \psi^f_{\text{test},2}])\)
  \STATE \(\mc{P}^i_{\varphi_{\text{test},1}} \leftarrow\) Partial order\((G_{\varphi_{\text{test},1}}, \psi^f_{\text{test},1})\)
  \STATE \(\mc{P}^k_{\varphi_{\text{test},2}} \leftarrow\) Partial order\((G_{\varphi_{\text{test},2}}, \psi^f_{\text{test},2})\) 
  
  \STATE \(E^r_{\varphi_{\text{test},1} \vee \varphi_{\text{test},2}} \subseteq E\) Deleting outgoing edges from \(\mc{V}^i_{0}\cup\mc{V}^k_{0} \subseteq V\) within \(G_{\varphi_{\text{test},1} \vee \varphi_{\text{test},2}}\)
  \STATE \(E^a_{\varphi_{\text{test},1} \vee \varphi_{\text{test},2}}\) Adding edges from \(\mc{V}^i_{0}\cup\mc{V}^k_{0} \subseteq V\) to subgraphs \(G_{\varphi_{\text{test},1}}\) and \(G_{\varphi_{\text{test},2}}\)
  \STATE \(E^r_{\varphi_{\text{test},1}} \subseteq E_{1}\) Deleting outgoing edges from \(\mc{V}^i_{0} \subseteq V_{1}\) within \(G_{\varphi_{\text{test},1}}\)
  \STATE \(E^a_{\varphi_{\text{test},1}}\) Adding edges from \(\mc{V}^i_{0} \subseteq V_{1}\) to subgraph \(G_{\varphi_{\text{test},1}\vee\varphi_{\text{test},2}}\)
  \STATE \(E^r_{\varphi_{\text{test},2}} \subseteq E_{2}\) Deleting outgoing edges from \(\mc{V}^k_{0} \subseteq V_{2}\) within \(G_{\varphi_{\text{test},2}}\)
  \STATE \(E^a_{\varphi_{\text{test},2}}\) Adding edges from \(\mc{V}^k_{0} \subseteq V_{2}\) to subgraph \(G_{\varphi_{\text{test},1}\vee\varphi_{\text{test},2}}\)
 
\STATE \(V_\text{aux} = V \cup V_{1} \cup V_{2}\)
\STATE \(E_\text{aux} = \big(E \setminus E^r_{\varphi_{\text{test},1} \vee \varphi_{\text{test},2}}\big)\cup \big(E_{1} \setminus E^r_{\varphi_{\text{test},2}} \big) \cup \big(E_{2} \setminus E^r_{\varphi_{\text{test},2}}\big) \cup E^a_{\varphi_{\text{test},2}} \cup E^a_{\varphi_{\text{test},1}} \cup E^a_{\varphi_{\text{test},1} \vee \varphi_{\text{test},2}}\)
\STATE \(G_\text{aux}=(V_\text{aux}, E_\text{aux})\)
 \RETURN $G_\text{aux}, \mc{P}^i_{\varphi_{\text{test},1} \vee \varphi_{\text{test},2}}, \mc{P}^k_{\varphi_{\text{test},1} \vee \varphi_{\text{test},2}}, \mc{P}^i_{\varphi_{\text{test},1}}, \mc{P}^k_{\varphi_{\text{test},2}}$ 
 \end{algorithmic} 
 \end{algorithm}
 \subsection{Live Lock}
\begin{figure}%
    \centering
    \subfloat[\centering
    \label{fig:rhws}]{{\includegraphics[width=0.45\textwidth]{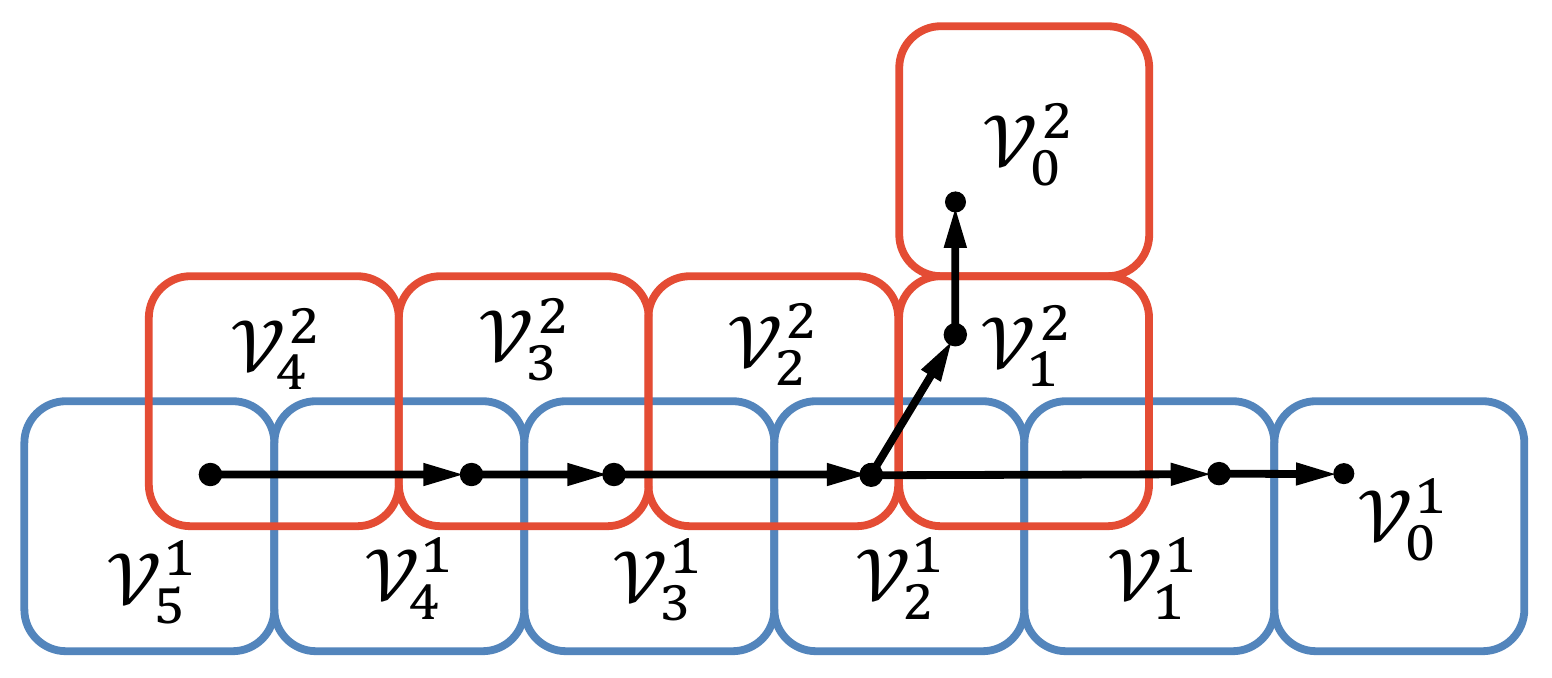}}} %
    \qquad
    \subfloat[\centering \label{fig:rhws_loop}]{{\includegraphics[width=0.45\textwidth]{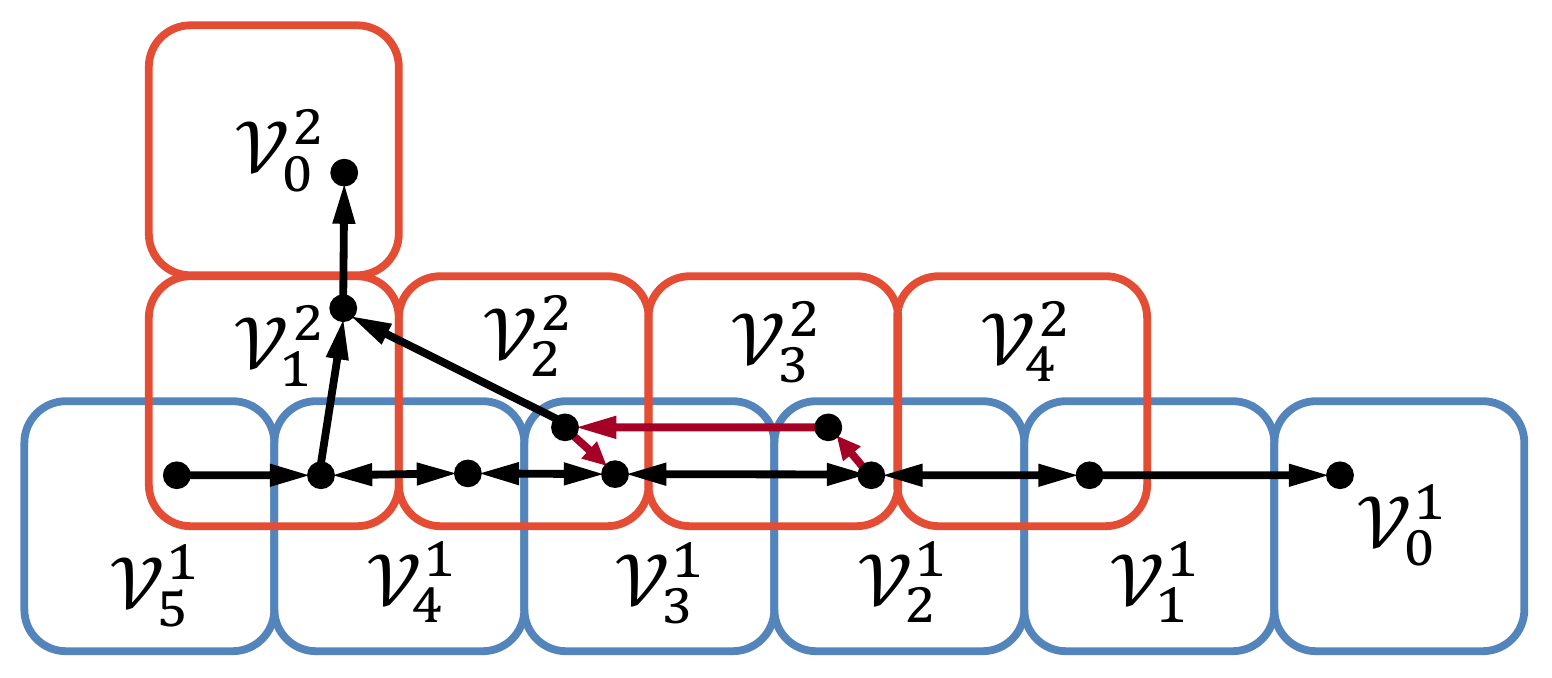}}} %
    \caption{Sketch of receding horizon winning set with and without cycle.}%
\end{figure}

 Depending on the construction of the partial order, the test could end up in a live lock. This is a result of planning over a short horizon for a disjunction of specifications, \(\psi^i_j\), each of which specifies progress on different partial orders. An example of naively applying \(\mc{W}^{\mc{I}}\) as a filter is given in Figure~\ref{fig:rhws_loop}, where an execution can get stuck in the loop (\(\mathcal{V}^1_2 \rightarrow \mathcal{V}^2_3 \rightarrow \mathcal{V}^2_2 \rightarrow \mathcal{V}^1_3 \rightarrow \ldots\)), where progress towards goals \(1\) and \(2\) happens infinitely often but neither of the goals are reached. Consider the example of a roundabout, where the system always makes progress towards one of the exits while driving around the roundabout, even if it never chooses to take an exit.
 To address this, we propose removing a goal from \(\mathcal{I}\) that the test execution has stopped making progress towards, and store it in \(\mathcal{I}'\). If \(\mathcal{I}\) becomes empty before one of the goals are reached, we reset \(\mathcal{I}\) to have all goals stored in \(\mathcal{I}'\). 
\begin{remark}
This approach to ensuring that the test execution reaches one of the goals \(i \in \mc{I}\) requires that eventually, there exists a path.
\end{remark}

 \subsection{Example: Lane Change}
\begin{figure}%
    \centering
    \subfloat[\centering
    \label{fig:mcts5}]{{\includegraphics[width=0.45\textwidth]{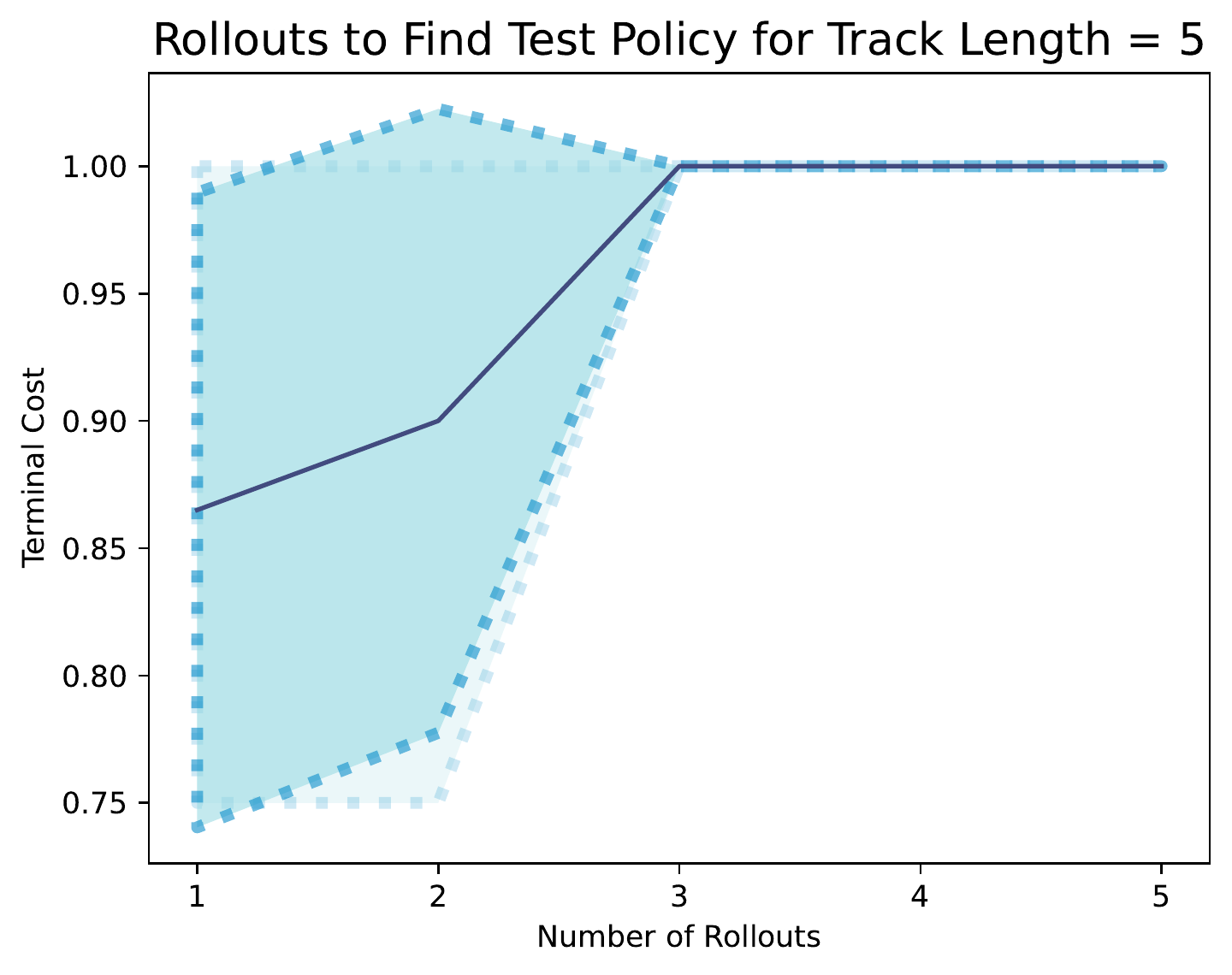}}} %
    \qquad
    \subfloat[\centering \label{fig:mcts10}]{{\includegraphics[width=0.45\textwidth]{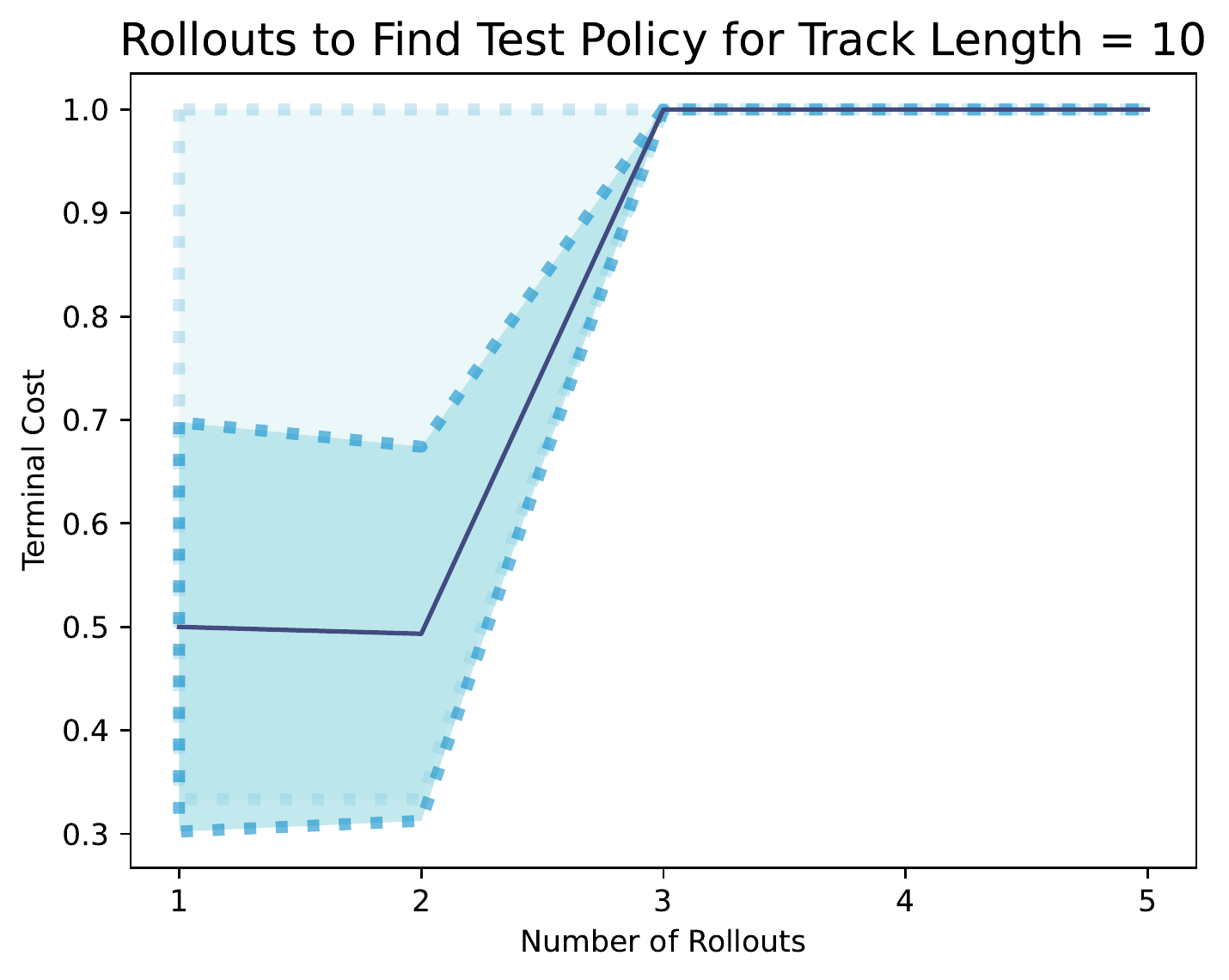}}} %
    \\
    \subfloat[\centering \label{fig:mcts15}]{{\includegraphics[width=0.45\textwidth]{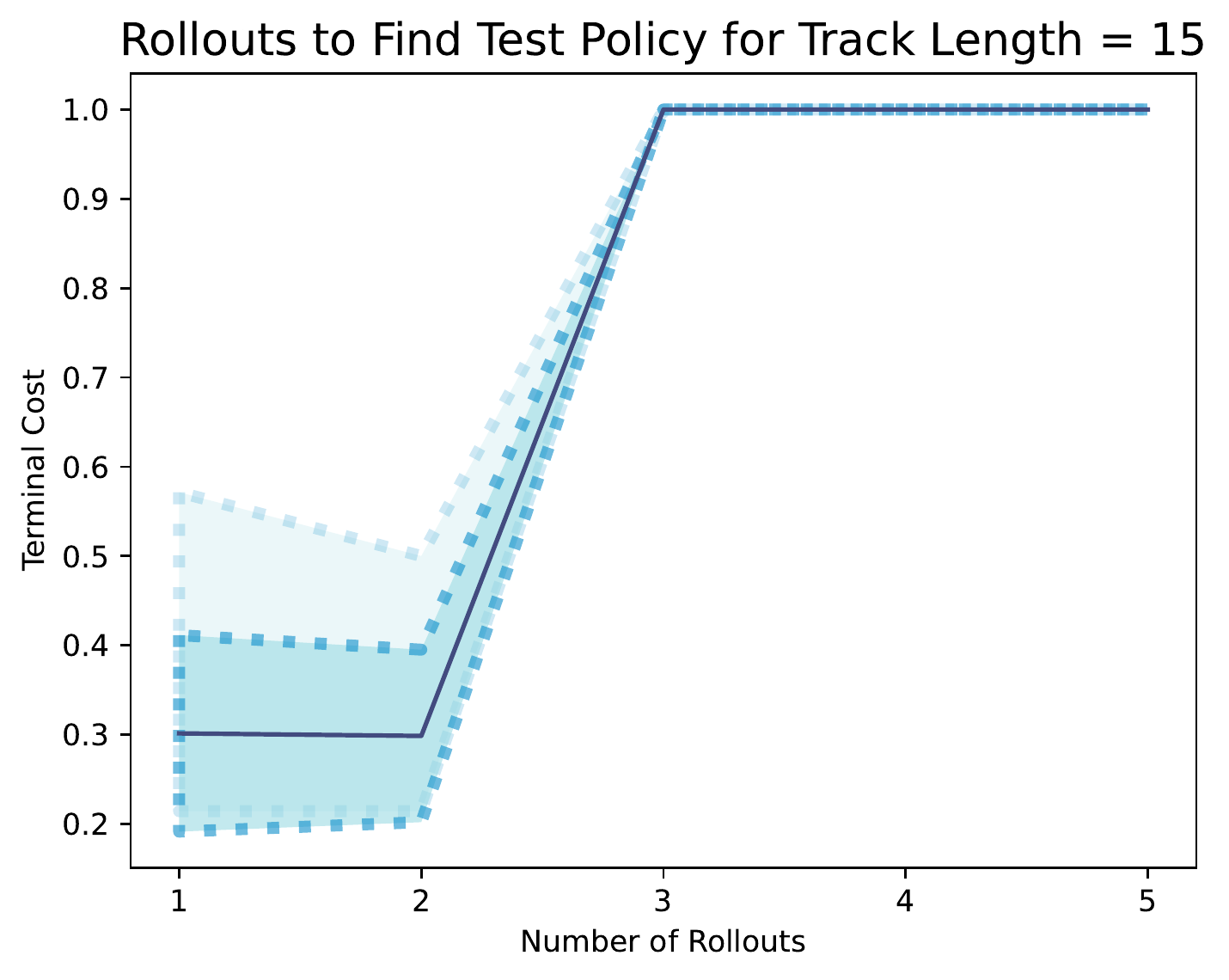}}} %
    
    \caption{The normalized mean terminal cost of the test execution found by our framework shown for a different number of rollouts for the track lengths 5, 10, and 15. The shaded areas represent the minimum and maximum value (light blue) and the standard deviation (blue) over 50 runs.}\label{fig:mcts_cost}
\end{figure}

\begin{figure}
    \centering
    \includegraphics[width=0.6\textwidth]{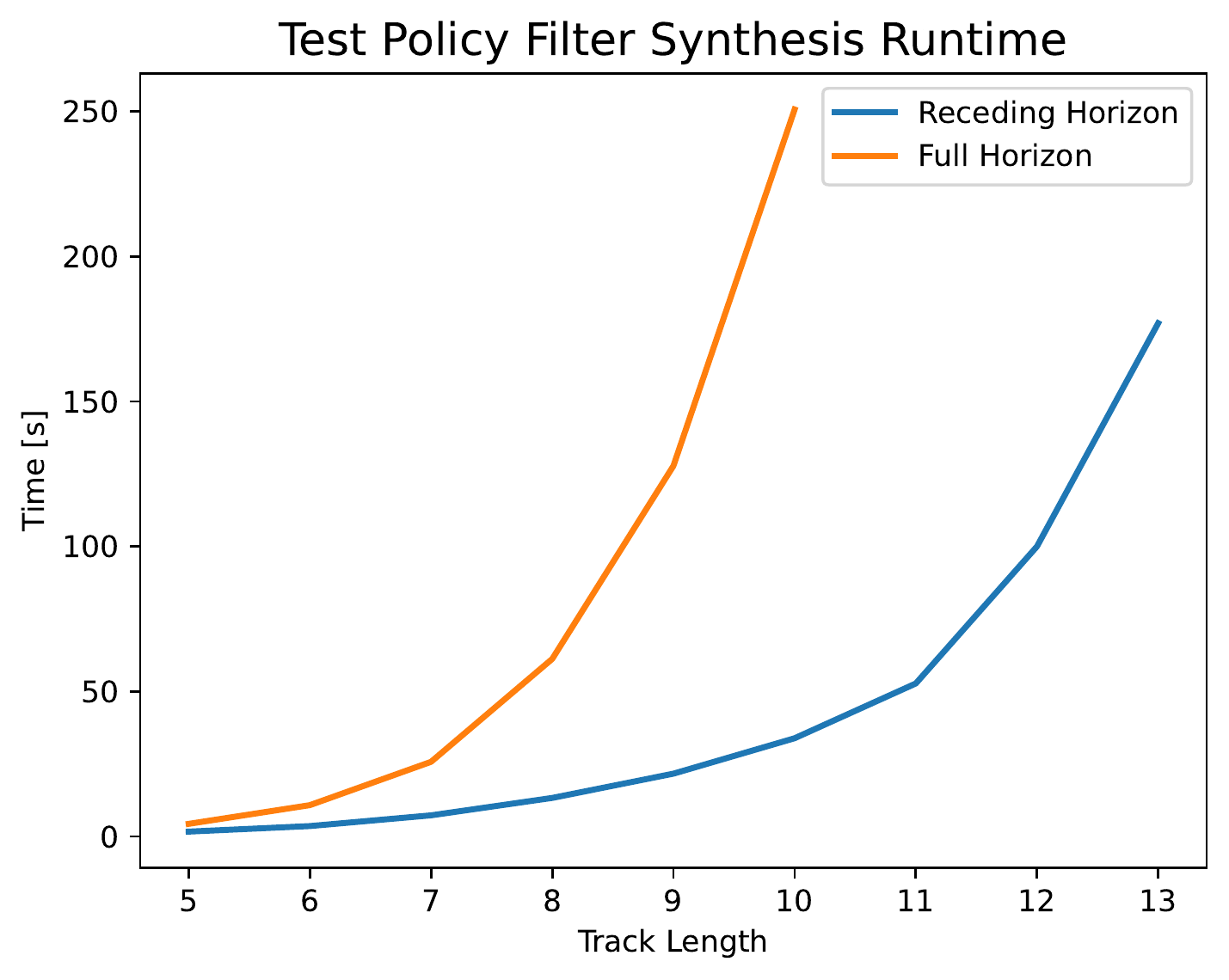}
    \caption{The computation time required to generate the winning set filter with the receding horizon approach and by computing the entire winning set for each possible goal at once. The experiments were run on a MacBook Pro with a 2.3 GHz Quad-Core Intel Core i7 processor with 32 GB RAM.}
    \label{fig:synth_time}
\end{figure}

On the lane change example, we analyzed the convergence of MCTS as the search procedure.
Figure~\ref{fig:mcts_cost} shows that the terminal cost (robustness metric) reaches the maximum value with a relatively low number of rollouts. This is due to the fact that we are applying our framework to a problem with a relatively small action space for the test environment, using the test policy filter, and MCTS as an online policy. Even though the state space of the lane change example grows significantly with an increase of the track length, the actions that the testers can take are at maximum four (both move, both stay, one moves / one stays). With the use of the winning set and depending on the positions of the system and testers, the number of possible actions can be smaller. Because only actions that remain in the winning set for the specification can be chosen, the search procedure quickly finds a policy that maximizes the cost. The number of iterations used by the online MCTS depends on the actions of the system and is upper bounded by the maximum duration of the test. As we find a search procedure online, every time that the test environment has to take its turn, MCTS executes the specified number of rollouts to choose the next action, and this continues until the test is finished.
In Figure~\ref{fig:synth_time} the runtime for the winning set synthesis is shown. We compare the runtime of the receding horizon approach to the synthesis of the full horizon winning set for each goal location at once. While the runtimes for both approaches increase significantly, the full horizon approach is already unable to generate a winning set for a track length of 11 for the same specifications. 

 \subsection{Example: Unprotected Left Turn}
  \begin{figure}
    \centering
    \includegraphics[width=0.5\textwidth]{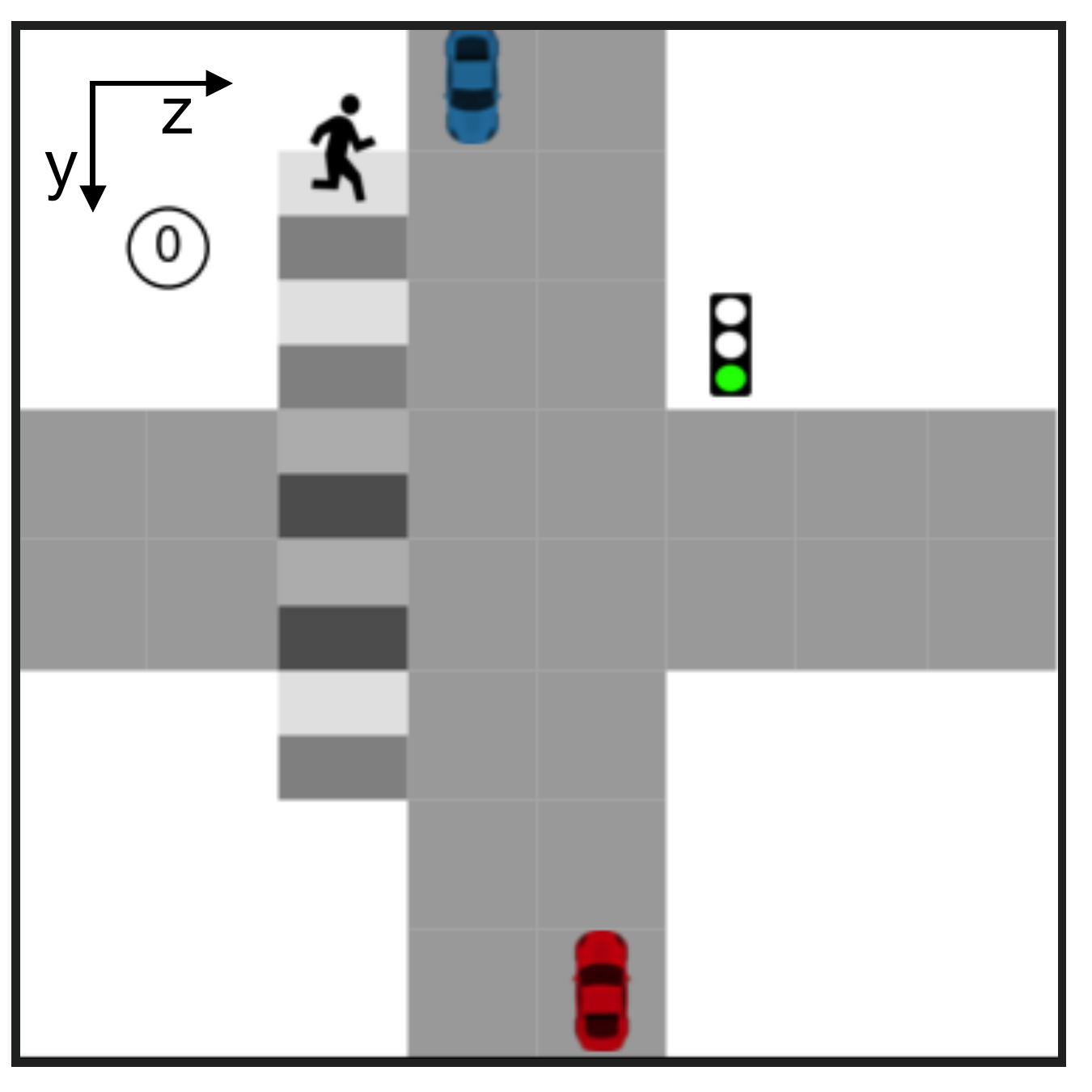}
    \caption{Layout of the unprotected left turn at intersection example. The system starts in cell (7,4) and wants to reach the goal cell (0,3), while the initial positions of the test agents are at the beginning of the road and crosswalk.}
    \label{fig:intersection_layout}
\end{figure}

The test specification for waiting for the test car is specified according to equation~(\ref{eq:test_spec}), with
\begin{equation}
\begin{split}
    \varphi^{init}_{sys} = (\textbf{x}_S \in I_S), \quad \square \lozenge \varphi_{sys}^f = \lozenge (\textbf{x}_S \in \mathcal{S}_G), \quad
     \square \lozenge \psi_{\text{test}}^f = \lozenge (\textbf{x}_S \in \mathcal{S}_C \land \textbf{x}_C \in T_C)\,,
\end{split}
\end{equation}
where the subscript $C$ denotes the tester car.
 In Figure~\ref{fig:intersection_layout}, the conventions used for the left turn at intersection example are depicted. The coordinate system starts in the upper left corner with cell $(y,z) = (0,0)$ and the $y$-axis facing south and the $z$-axis facing east. The crosswalk locations are numbered from north to south, starting with $0$.
 The initial states of the test agents are $\textbf{x}_C=(0,3)$ and $\textbf{x}_P=0$, and the initial state of the system is $\textbf{x}_S=(7,4)$. The goal state for the system is $\textbf{x}_G = (0,3)$. In this example $\textbf{x}_G$ is the only element in $\mathcal{S}_\mathcal{G}$.
 The states in which the system needs to wait for the pedestrian and the car, $\mathcal{S}_P$ and $\mathcal{S}_P$ respectively, are both $\textbf{x}=(4,4)$ for this layout. The states of the tester car, for which the system has to wait are given as $\mathcal{T}_C = \{(0,3), (1,3),(2,3),(3,3)\}$ and the states of the pedestrian, for which the system has to wait are $\mathcal{S}_P=\{1,2,3,4,5\}$, which represent the cells on the crosswalk, that map to grid coordinates. Note that if the pedestrian is in cell $0$, the system is not required to wait for the pedestrian, as she is too far away from the road.
 The traffic light sequence is predetermined, the light will be green for a fixed number of time steps $t_{g}$, followed by $t_y$ time steps of yellow and red for $t_r$ time steps. We are assuming that the system designer supplied the robustness metric as the time until the traffic light turns red, resulting in a harder test the closer the light is to red once the system successfully takes the turn.

\end{document}